\documentclass{emulateapj}
\usepackage{epsfig}
\usepackage{subfigure}
\usepackage[figuresright]{rotating}
\usepackage{natbib}
\usepackage{xcolor}
\usepackage[colorlinks, citecolor=blue]{hyperref}
\usepackage{longtable}
\usepackage{threeparttable}
\usepackage{url}
\usepackage{float}
\usepackage{amsmath}
\usepackage{makecell}

\shorttitle{Fallback accretion on millisecond magnetar}
\shortauthors{Yu et al.}
\slugcomment{}

\begin{document}
%\linenumbers

\title{Can fallback accretion on magnetar model power the X-ray flares simultaneously observed with 
gamma-rays of Gamma-ray bursts?}
\author{Wen-Yuan Yu\altaffilmark{1}, Hou-Jun L\"{u}\altaffilmark{1}, Xing Yang\altaffilmark{1}, Lin Lan
\altaffilmark{2}, Zhe Yang\altaffilmark{1}}\altaffiltext{1}{Guangxi Key Laboratory for Relativistic
Astrophysics, School of Physical Science and Technology, Guangxi University, Nanning 530004, China;
lhj@gxu.edu.edu} \altaffiltext{2}{CAS Key Laboratory of Space Astronomy and Technology, National
Astronomical Observatories, Chinese Academy of Sciences, China}

\begin{abstract}	
%\linenumbers
The prompt emission, X-ray plateau, and X-ray flares of Gamma-ray bursts (GRB) are thought to be
from internal dissipation, and the magnetar as the central engine with propeller fallback accretion
is proposed to interpret the observed phenomena of GRBs. In this paper, by systematically searching
for X-ray emission observed by Swift/Xry Telescope, we find that seven robust GRBs include both
X-ray flares and plateau emissions with measured redshift. More interestingly, the X-ray
flares/bumps for those seven GRBs are simultaneously observed in the gamma-ray band. By adopting
the propeller fallback accretion model to fit the observed data, it is found that the free
parameters of two GRBs (140512A and 180329B) can be constrained very well, while in the other five
cases, more or less, they are not all sufficiently constrained. On the other hand, this requires
that the conversion efficiency of the propeller is to be two or three times higher than that of the
spindown dipole radiation of the magnetar. If this is the case, it is contradictory to the
expectation from the propeller model: namely, a dirtier ejecta should be less efficient in
producing gamma-ray emissions. Our results hint that at least the magnetar central engine with
propeller fallback accretion model cannot interpret very well both the GRB X-ray flares
simultaneously observed in the gamma-ray band and the X-ray flares of GRBs with a high Lorentz
factor.
\end{abstract}


\section{Introduction}
%\linenumbers
The majority of long-duration gamma-ray bursts (LGRBs) are thought to originate from the deaths of
massive stars, excepting for several specific events, i.e., GRB 060614 \citep{2006Natur.444.1044G},
GRB 191019A \citep{2023NatAs...7..976L}, GRB 211211A
\citep{2022Natur.612..223R,2022Natur.612..228T,2022Natur.612..232Y}, and GRB 211227A
\citep{2022ApJ...931L..23L}, and the $``$collapsar$"$ model has been widely accepted as the
standard paradigm for LGRBs \citep{1993ApJ...405..273W,1999ApJ...524..262M}. A black hole or a
rapidly spinning, strongly magnetized neutron star (called a millisecond magnetar) as the central
engine of a GRB may be formed after collapse
\citep{1992Natur.357..472U,1994MNRAS.270..480T,1998PhRvL..81.4301D,1999ApJ...518..356P,2000ApJ...537..810W,2001ApJ...552L..35Z,2008MNRAS.385.1455M,2010MNRAS.402..705L,2013ApJ...765..125L,2014ApJ...785...74L,2017NewAR..79....1L}.
It successfully launches the relativistic outflow, to break out of the stellar envelope of the
progenitor star, and produces prompt gamma-ray emission, which passes through the photosphere,
internal shocks, or undergoes magnetic dissipation, becoming transparent
\citep{1993ApJ...405..278M,1993MNRAS.263..861P,1994ApJ...430L..93R,2011ApJ...726...90Z,2018pgrb.book.....Z}.
The broadband afterglow emission is attributed to synchrotron emission from the external shock when
the outflow is decelerated by a circumburst medium
\citep{1997ApJ...476..232M,1998ApJ...497L..17S,2016ApJ...816...72Z}.

There is a small fraction of LGRBs in which the prompt emission shows two or more sub-bursts of
emission in the light curves, with quiescent times of up to hundreds of seconds, detected by both
Fermi/Gamma-ray Burst Monitor (GBM) and Swift/Burst Alert Telescope (BAT)
\citep{1995ApJ...452..145K,2005MNRAS.357..722L,2008ApJ...685L..19B,2013ApJ...775...67B,2014ApJ...789..145H,2018ApJ...862..155L}.
\cite{2018ApJ...862..155L} investigated the spectral and temporal properties of two sub-bursts of
emission in the prompt emission, and they did not find any statistically significant correlation in
the duration and peak energy ($E_{\rm p}$) of the two sub-bursts of emission. This suggests that
those two or more sub-bursts of emission of GRB are likely to be from the same origin.

Thanks to Swift, a multiwavelength GRB mission \citep{2004ApJ...611.1005G} has led to great
progress in understanding the nature of this phenomenon (review for \citealt{2007ChJAA...7....1Z}).
Especially, the prompt slewing capability of the onboard X-ray Telescope (XRT;
\citealt{2004SPIE.5165..201B}) has revealed the discovery of a canonical X-ray light curve
following the prompt emission, which shows successively four power-law decay segments (i.e., an
initial steep decay segment, a shallow decay segment, a normal decay segment, and a post-jet-break
phase), with superimposed erratic flares
\citep{2006ApJ...642..389N,2006ApJ...647.1213O,2006ApJ...642..354Z}. Moreover, a small fraction of
GRBs show an internal plateau\footnote{An internal plateau with a near flat light curve, before
rapidly falling off with a decay index $\alpha >3$. Throughout the paper, the convention $F_\nu
\propto t^{-\alpha} \nu^{-\beta}$ is adopted.} in the X-ray light curve
\citep{2007ApJ...662.1111L,2007ApJ...665..599T,2010MNRAS.402..705L,2010MNRAS.409..531R,2013MNRAS.430.1061R,2014ApJ...785...74L}.

From the theoretical point of view, both the internal plateau and the X-ray flares are inconsistent
with any external shock model, but must be attributed to the internal dissipation of a central
engine wind. There are two forms of continuous energy injection into the external forward shock.
One involves invoking a long-lasting central engine such as a spinning-down millisecond magnetar
\citep{1998PhRvL..81.4301D,2001ApJ...552L..35Z}. The other one involves a stratification of the
ejecta Lorentz factor in an impulsively ejected fireball
\citep{1998ApJ...496L...1R,2000ApJ...535L..33S,2012ApJ...761..147U}. The widely discussed model for
dealing with the energy injection process involves assuming a millisecond magnetar as the central
engine
\citep{2001ApJ...552L..35Z,2007ApJ...659.1420T,2007ApJ...665..599T,2014ApJ...785...74L,2015ApJ...813...92R,2018ApJ...869..155S,2021ApJ...918...12F}.
Within the scenario of a magnetar central engine, the energy injection to explain the plateau (or
shallow decay) phase is from the dipole radiation of the magnetar spindown. This has previously
been studied for LGRBs (e.g., \citealt{2010MNRAS.402..705L,2011A&A...526A.121D,
2014ApJ...785...74L}), short GRBs (e.g.,
\citealt{2013MNRAS.430.1061R,2015ApJ...805...89L,2017ApJ...849...71L}), as well as the extended
emission of short GRBs (e.g., \citealt{2013MNRAS.431.1745G}). These works assumed a constant rate
of spindown and therefore a constant level of dipole luminosity.

Another component, a flare/bump in the X-ray emission, is also of an important clue to
understanding the physical process in the framework of the magnetar central engine. This has been
discovered in a good fraction of GRBs
\citep{2005Natur.438..994B,2007ApJ...671.1903C,2007ApJ...671.1921F,2010A&A...524A..64C,2010MNRAS.406.2149M}.
In most cases, flares/bumps are superposed on either the steep decay phase or the shallow decay
phase \citep{2006ApJ...642..389N,2006ApJ...642..354Z}. \cite{2014ApJ...795..155P} found that a good
fraction of GRB X-ray flares, which were observed simultaneously by both BAT and XRT on board the
Swift mission, obeyed the same power-law spectral fit. By studying the properties of both temporal
and spectral behaviors, this suggests that the X-ray flares are produced by late central engine
activities, and may share the same physical origin as the prompt emission of GRBs
\citep{2005Sci...309.1833B,2016ApJS..224...20Y}. If this is the case, a small fraction of the GRB
X-ray flares at least can be used to constrain the Lorentz factors, which range from a few tens to
hundreds \citep{2010ApJ...724..861J,2015ApJ...807...92Y}. On the other hand, several models have
also been invoked to interpret the X-ray flares, such as the tail of prompt emission with a
high-latitude curvature effect \citep{2006ApJ...646..351L}, delayed magnetic dissipation activity
as the ejecta decelerates \citep{2006A&A...455L...5G}, and anisotropic emission in the blast wave
comoving frame \citep{2011MNRAS.410.2422B}. Moreover, the magnetar central engine with the delayed
onset of a propeller regime, which accelerates local material via magnetocentrifugal slinging, has
also been proposed to explain the X-ray flares in GRBs
\citep{2017MNRAS.470.4925G,2018MNRAS.478.4323G}. This model has also been invoked to account for
supernova explosions \citep{2011ApJ...736..108P} and X-ray plateaus for short GRBs
\citep{2014MNRAS.438..240G}. However, the main issue of the propeller mechanism for explaining the
prompt emission of GRBs is baryon loading. The propeller mechanism expels a lot of unaccreted
material from the system, and the outflow must be very dirty and nonrelativistic if the material is
expelled in the funnel region. This means that no matter how large the power is, it cannot power a
GRB itself.

One interesting question is whether the propeller mechanism with fallback accretion on the
millisecond magnetar model can indeed interpret an X-ray light curve of a GRB that is composed of
both plateau emission and an X-ray flare/bump. In this paper, by systematically searching for X-ray
emission observed by Swift/XRT, we find X-ray light curves of seven GRBs composed of plateau
emission and X-ray flares with measured redshift. The basic propeller with fallback accretion model
of the magnetar central engine is shown in Section 2. In Section 3, we present the criteria for the
sample selection, and the fitting results for those seven GRBs with the propeller model, by
adopting the method of Markov Chain Monte Carlo (MCMC, \citealt{2003itil.book.....M}) are reported
in Section 4. Finally, we summarize our discussion and conclusion in Section 5. Throughout the
paper, a concordance cosmology with the parameters $H_0=71~\rm km~s^{-1}~Mpc^{-1}$,
$\Omega_M=0.30$, and $\Omega_{\Lambda}=0.70$ is adopted.

%%%%%%%%%%%%%%%%%%%%%%%%%%%%%%%%%%%%%%%%%%%%%%%%%%%%%%%%%%%%%%%%%%%%%%%%%%%%%%%%%%%%%
\section{The magnetar propeller with fallback accretion model}
Previously, the propeller with fallback accretion model of a magnetar central engine was invoked to
explain the short GRB with extended emission
\citep{2014MNRAS.438..240G,2017MNRAS.470.4925G,2020MNRAS.492.3622L}, X-ray plateau emission and
X-ray flares in long-duration GRBs \citep{2018MNRAS.478.4323G}, as well as some stripped-envelope
Supernovae \citep{2021ApJ...914L...2L}. In this section, we will briefly introduce the model of
magnetar propeller with fallback accretion.

The propeller regime is basically defined by the relationship between the Alfv\'en radius ($r_{\rm
m}$) and the co-rotation radius ($r_{\rm c}$). Generally speaking, when $r_{\rm c}<r_{\rm m}$, the
materials already within $r_{\rm c}$ accrete to the surface of the magnetar, while the materials
within the range of $r_{\rm c}$ and $r_{\rm m}$ are propelled away at $r_{\rm m}$. If the power of
propeller is not strong enough for materials, it cannot reach the potential well. Then, the
materials will return to the disc without any emission to be detected (called ``propeller regime"
\citealt{1975A&A....39..185I}). Based on the paper of \cite{2017MNRAS.470.4925G}, $r_{\rm m}$ and
$r_{\rm c}$ can be defined as follows,
\begin{equation}
r_{\mathrm{m}}=\mu^{4/7}(GM)^{-1/7}\left(\dfrac{3M_{\mathrm{d}}}{t_{\nu}}\right)^{-2/7},
\end{equation}
\begin{equation}
    r_\mathrm{c}=(GM/\omega^2)^{1/3},
\end{equation}
where $G$ is the gravitational constant, $\mu=B_{\rm p}R^3$ is the magnetic dipole moment, $B_{\rm
p}$ is the surface magnetic field, $M$ and $R$ are the mass and radius of magnetar, respectively.
$t_\nu=R_{\rm d}/ \alpha c_{\rm s}$ is the viscous time-scale, $\alpha$ and $c_{\rm s}$ are the
viscosity prescription and the speed of sound in the disc, respectively. $\omega$ is the angular
frequency of the magnetar, $M_{\rm d}$ and $R_{\rm d}$ are the mass and radius of disc,
respectively. In our calculations, we adopt $\alpha = 0.1$ and $c_{\rm s}=10^{7}\mathrm{cm} \cdot
\mathrm{s}^{-1}$ \citep{2014MNRAS.438..240G}.

At the Alfv\'en radius $r_{\rm m}$, the effect of materials from the accretion is smaller than that of
magnetic field of magnetar. While, at the co-rotation radius $r_{\rm c}$, the rotation rate of matter in the
disc is the same as that of the stellar surface. Since $r_{\rm m}$ and $r_{\rm c}$ depend on the disc mass
and the frequency of the magnetar, respectively. The evolution of disc mass and frequency with time can be
expressed as follows \citep{2017MNRAS.470.4925G},
\begin{equation}\label{daf}
    \dot M_\mathrm{d}=\dot M_\mathrm{fb}-\dot M_\mathrm{prop}-\dot M_\mathrm{acc},
\end{equation}
\begin{equation}\label{omiga}
    \dot{\omega}=\frac{N_{\mathrm{acc}}+N_{\mathrm{dip}}}{I}.
\end{equation}
Here, $\dot M_\mathrm{fb}$ is the fallback mass changed with time, $\dot M_\mathrm{prop}$ and $\dot
M_\mathrm{acc}$ are the mass lost via the propeller mechanism and the accretion to the magnetar,
respectively. $N_\mathrm{acc}$ and $N_\mathrm{dip}$ are the accretion and dipole torques acting on
the magnetar, respectively. $I=0.35M R^2$ is the magnetar moment of inertia.

Based on the description of \cite{2017MNRAS.470.4925G}, $\dot M_\mathrm{fb}$, $\dot M_\mathrm{prop}$, and
$\dot M_\mathrm{acc}$ can be defined as follows,
\begin{equation}
    \dot{M}_{\rm{fb}}=\frac{M_{\rm{fb}}}{t_{\rm{fb}}}\left(\dfrac{t+t_{\rm{fb}}}{t_{\rm{fb}}}\right)^{-5/3},
\end{equation}
\begin{equation}
    \dot M_{\rm{prop}}=\eta\left(\dfrac{M_{\rm{d}}}{t_\nu}\right),
\end{equation}
\begin{equation}
    \dot{M}_{\rm{acc}}=(1-\eta)\left(\dfrac{M_{\rm{d}}}{t_\nu}\right),
\end{equation}
where $M_{\rm{fb}}=\delta M_{\rm d,0}$ is the available fallback mass, $M_{\rm d,0}$ is the initial
mass of disc, and $\delta$ is the ratio between fallback mass and initial disc mass.
$t_{\rm{fb}}=\epsilon t_{\nu}$ is the fallback time-scale, and $\epsilon$ is the ratio between
fallback time-scale and viscous time-scale. $\eta=\dfrac{1}{2}(1+\tanh[n(\Omega-1)])$ is the
efficiency of the propeller mechanism, and it related to the "fastness parameter" $\Omega=\omega/(G
M/r_{\mathrm{m}}^{3})^{1/2}=(r_{\mathrm{m}}/r_{\mathrm{c}})^{3/2}$. In this work, we adopt $n=1$ to
do the calculations.

For the dipole torque, we adopt the classical solution as given by \citep{1983bhwd.book.....S} and
\citep{2011ApJ...736..108P},
\begin{equation}\label{Ndip}
    N_{\text{dip}}=-\dfrac{\mu^2\omega^3}{6c^3}.
\end{equation}
Moreover, $N_\mathrm{acc}$ has two forms that depend on the relationship between $r_{\rm m}$ and
$R$. It reads as
\begin{eqnarray}\label{Nacc1}
    N_{\mathrm{acc}}=(GMr_{\mathrm{m}})^{1/2}(\dot{M}_{\mathrm{acc}}-\dot{M}_{\mathrm{prop}}), ~ r_{\rm m}>R
\end{eqnarray}
\begin{eqnarray}\label{Nacc2}
    N_{\mathrm{acc}}=(GMR)^{1/2}(\dot{M}_{\mathrm{acc}}-\dot{M}_{\mathrm{prop}}), ~
 r_{\rm m}<R
\end{eqnarray}

Then, do the integration of equations \ref{daf} and \ref{omiga}, one can estimate the components of
propelled luminosity and dipole luminosity,
\begin{equation}
    L_{\text{prop}}=\eta_{\text{prop}}\left[-\omega N_{\text{acc}}-\left(\eta
    \dfrac{GMM_{\text{d}}}{r_{\text{m}}t_{\nu}}\right)\right],
\end{equation}
\begin{equation}
    L_{\mathrm{dip}}=\eta_{\mathrm{dip}}\dfrac{\mu^2\omega^4}{6c^3}.
\end{equation}\par
Here, $\eta_\text{prop}$ and $\eta_\text{dip}$ are the conversion efficiencies of propeller and
dipole energy-luminosity, respectively. The total luminosity is therefore defined as follows,
\begin{equation}
    L_{\mathrm{tot}}=(1/f_{\mathrm{B}})(L_{\mathrm{dip}}+L_{\mathrm{prop}}),
\end{equation}
where $1/f_{\rm B}$ is the fraction of the stellar sphere that is emitting, and it is related to
the half-opening angle of the jet ($\theta_j$). One has
$f_{\mathrm{B}}=1-\operatorname{cos}\left(\theta_{\mathrm{j}}\right)$
\citep{1999ApJ...525..737R,1999ApJ...519L..17S}.

\begin{figure*}[h]
\centering
\includegraphics[angle=0,scale=0.6]{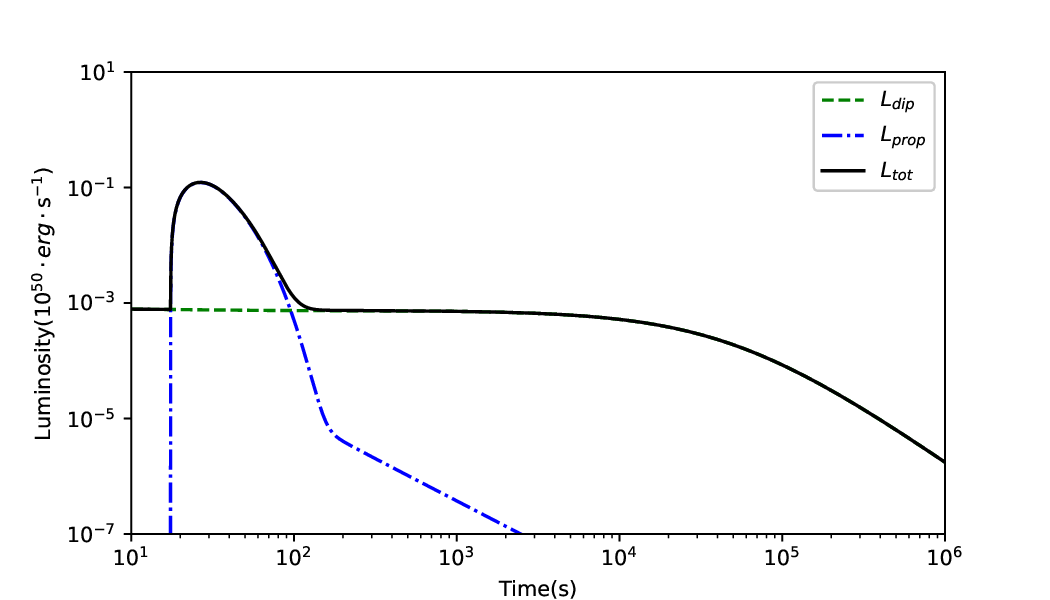}
\caption{X-ray light curve generated by propeller model with
fallback accretion (blue dash-dotted line) and
dipole radiation of spin-down (green dashed line) of
magnetar central engine. The black solid line is the sum
of propeller and dipole radiation of spin-down.}
\label{Fig1}
\end{figure*}

By given some typical parameters of magnetar propeller with fallback accretion model, such as $R =
10^{6}~\rm cm$, $M=1.4~M_{\odot}$, $B_{\rm p}=10^{15}~\rm G$, $P_0=10^{-3}~\rm s$, $M_{\rm d,0} =
10^{-3} M_{\odot}$, $R_{\rm d} = 10^7~\rm cm$, $\epsilon = 0.1$, $\delta = 10^{-4}$, $\eta_{\rm
prop}= 0.8$, $\eta_{\rm dip}= 0.1$, and $1/f_B = 10^{1.7}$, one can plot the picture of luminosity
as function of time for both propeller and dipole radiations (see Figure \ref{Fig1}).

%%%%%%%%%%%%%%%%%%%%%%%%%%%%%%%%%%%%%%%%%%%%%%%%%%%%%%%%%%%%%%%%%%%%%%%%%%%%%%%%%%%%%

\section{Sample Selection and Data Reduction}
The XRT data is downloaded from the Swift data archive and the UK Swift Science Data
Center\footnote {http://www.swift.ac.uk/burst\_analyser/}
\citep{2007A&A...469..379E,2009MNRAS.397.1177E}. Our entire sample includes more than 1718 GRBs
observed by Swift/XRT between January 2005 and June 2023. The magnetar signature typically exhibits
a shallow decay phase (or plateau) followed by a normal decay in X-ray emission when it is spinning
down, and the delayed onset of a propeller regime that accelerates local material via
magneto-centrifugal slinging can produce the X-ray flares. We only focus on the long-duration GRBs
with both flare/bump and plateau emissions observed in X-ray afterglow, and 154 GRBs are too faint
to be detected in the X-ray band, or do not have enough photons to extract a reasonable X-ray light
curve. Then, we select the GRBs whose X-ray emission exhibits the feature of plateau followed by a
normal decay, and adopt a smooth broken power-law function to fit \citep{2007ApJ...662.1111L},
\begin{eqnarray}
F_1(t)=F_{01}\left [
\left (\frac{t}{t_p}\right)^{\omega\alpha_1}+\left(
\frac{t}{t_p}\right)^{\omega\alpha_2}\right]^{-1/\omega},
\label{SBPL}
\end{eqnarray}
where fixed $\omega=3$ represents the sharpness of the peak and $t_p$, $\alpha_1$, and $\alpha_2$
are the peak time, and decay slope of plateau and normal decay phase, respectively.

Within the above scenario, three criteria are adopted for our sample selection. (i) We focus on
those long-duration GRBs that show such a transition from shallow decay to normal decay in the
X-ray light curves, but require that the decay slope of the normal decay segment following the
plateau phase should be in the range of $-1$ to $-2$ \citep{2018MNRAS.480.4402L}; (ii) It must
include the X-ray flare/bump in the X-ray light curve; (iii) In order to estimate the intrinsic
luminosity of the plateau emission, the redshift needs to be measured. By adopting the criteria for
our sample selection, only seven robust cases identified to satisfy the above three criteria. The
X-ray light curves are shown in Figure \ref{Fig3} (also see Table 1 for a summarized).

\begin{figure*}[h]
\centering
\includegraphics[angle=0,scale=0.35]{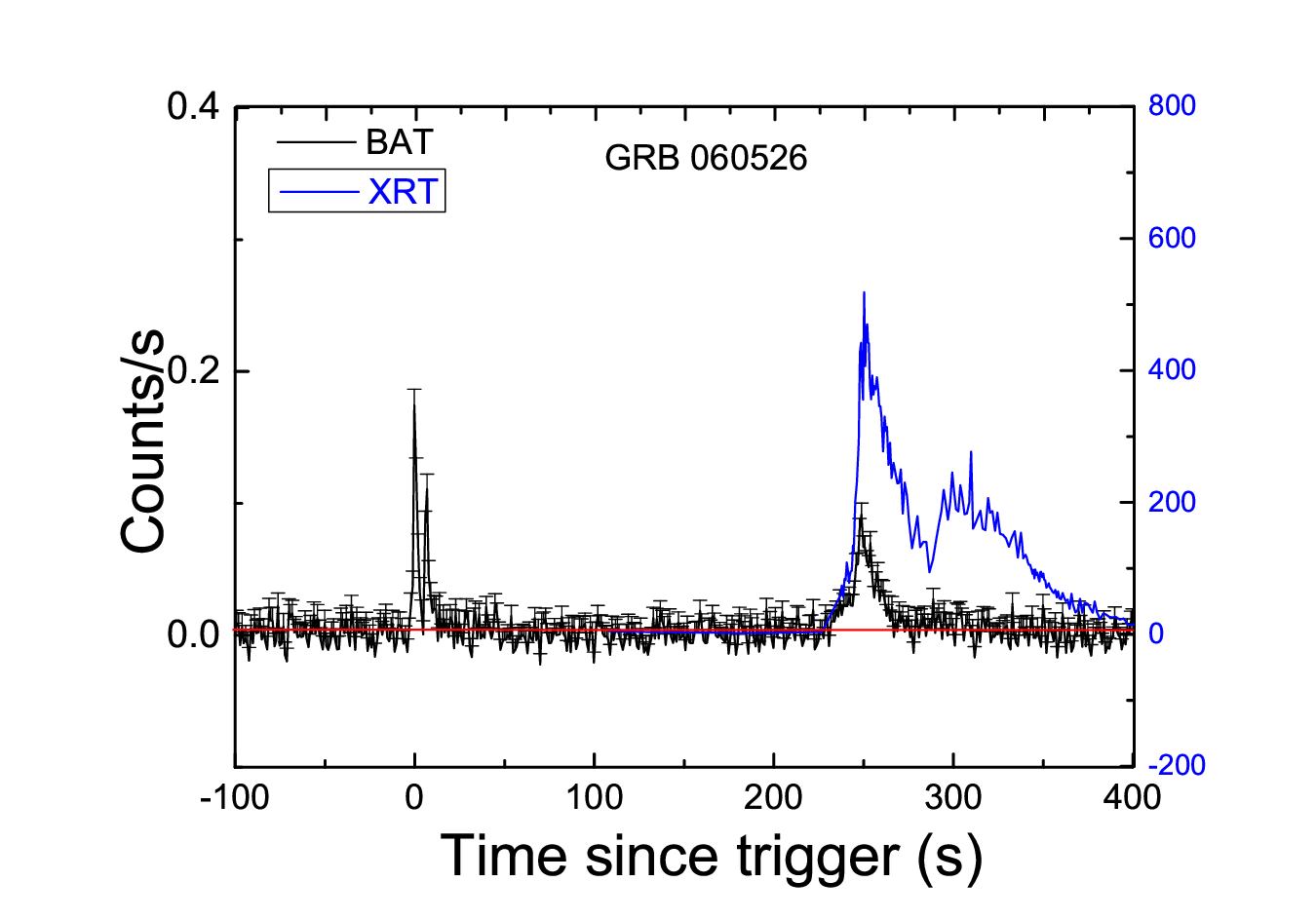}
\includegraphics[angle=0,scale=0.35]{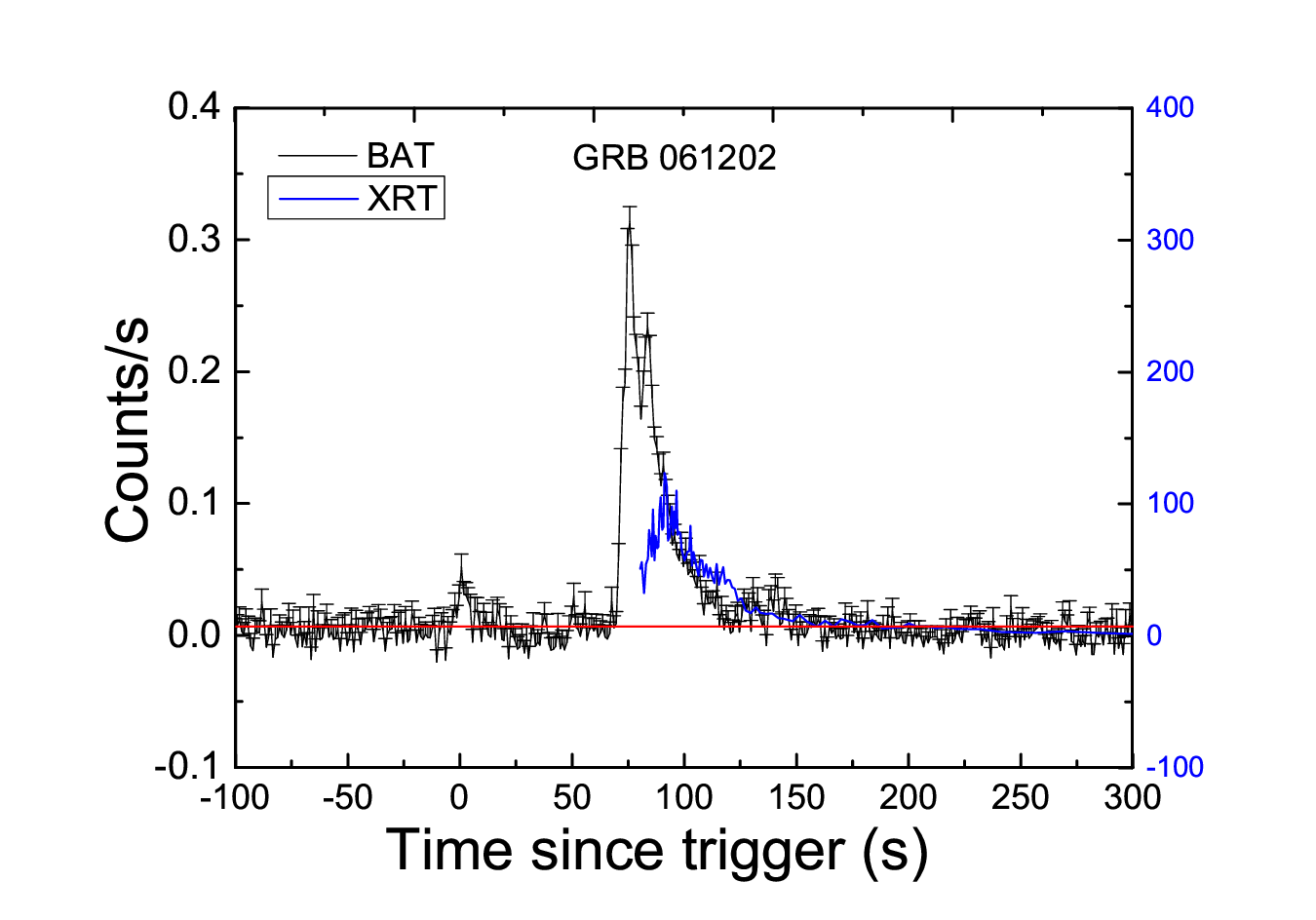}
\includegraphics[angle=0,scale=0.35]{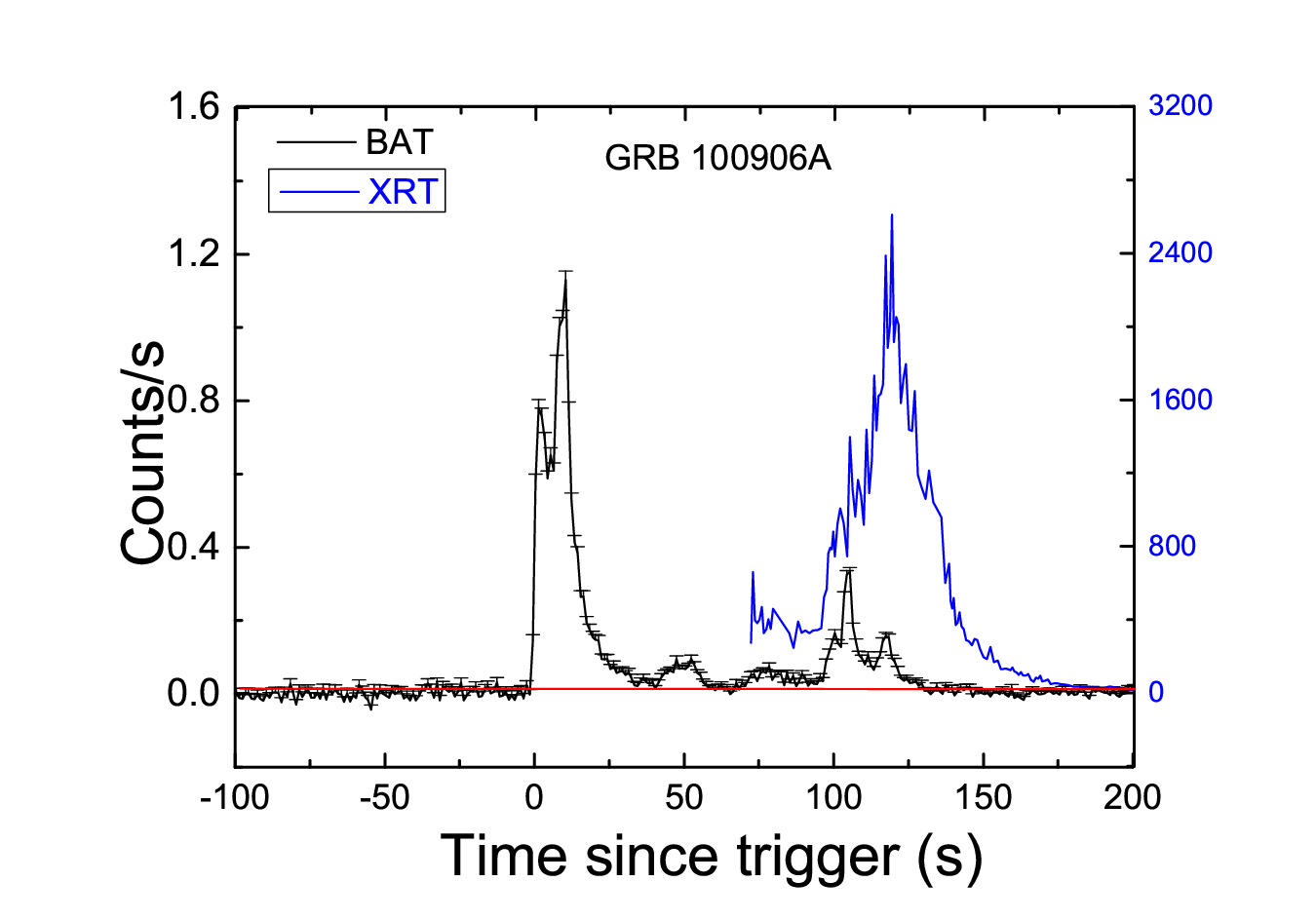}
\includegraphics[angle=0,scale=0.35]{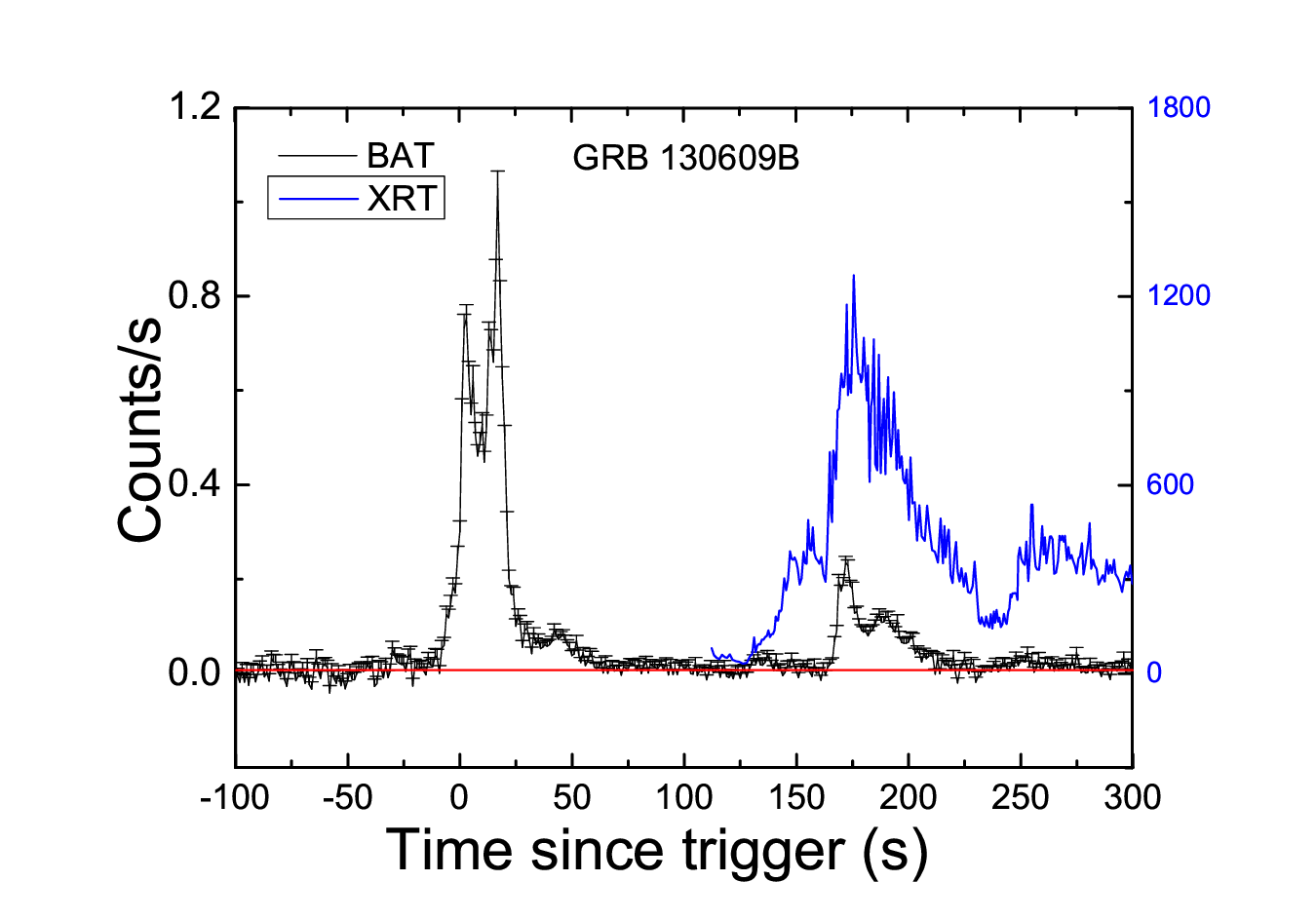}
\includegraphics[angle=0,scale=0.35]{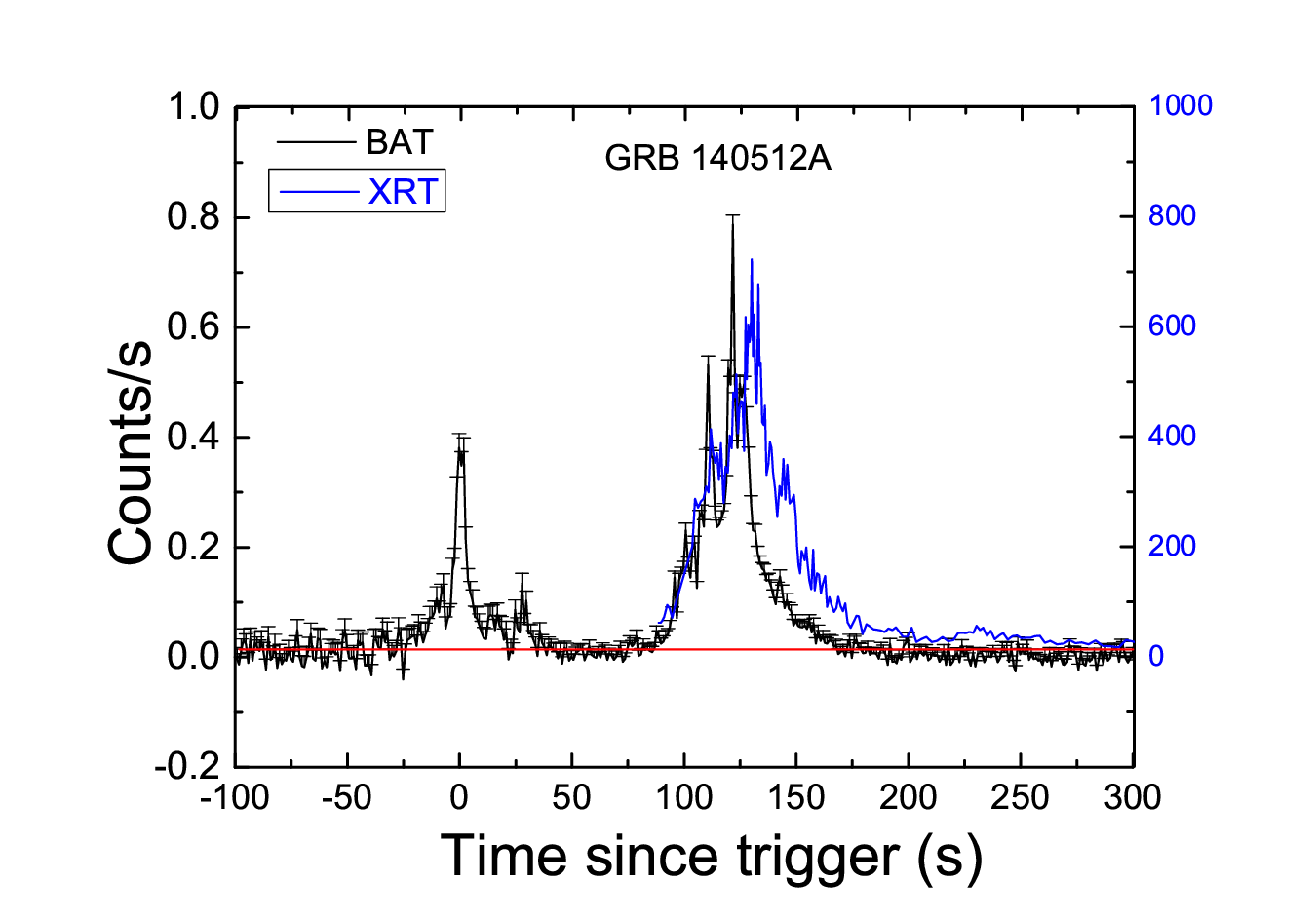}
\includegraphics[angle=0,scale=0.35]{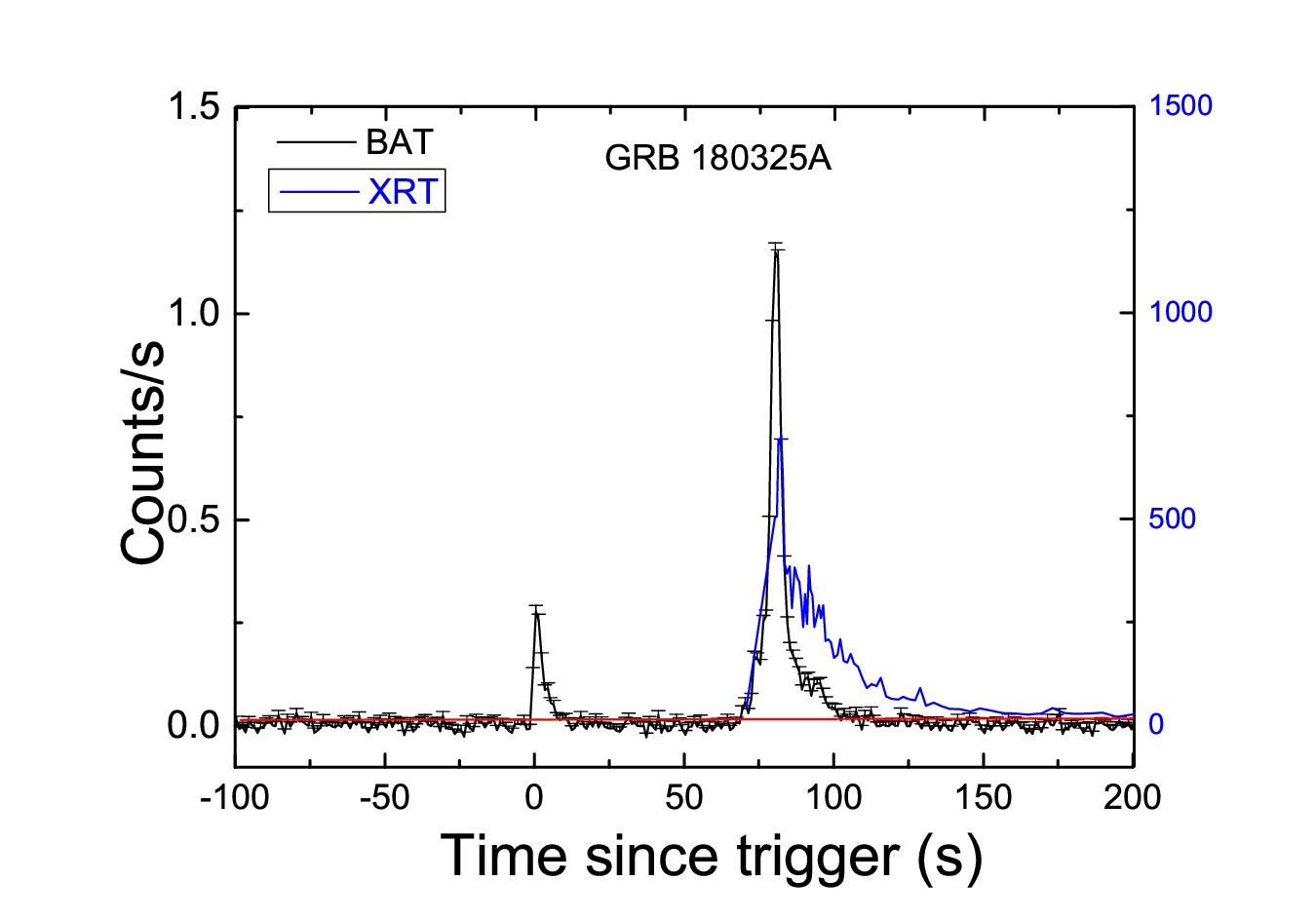}
\includegraphics[angle=0,scale=0.35]{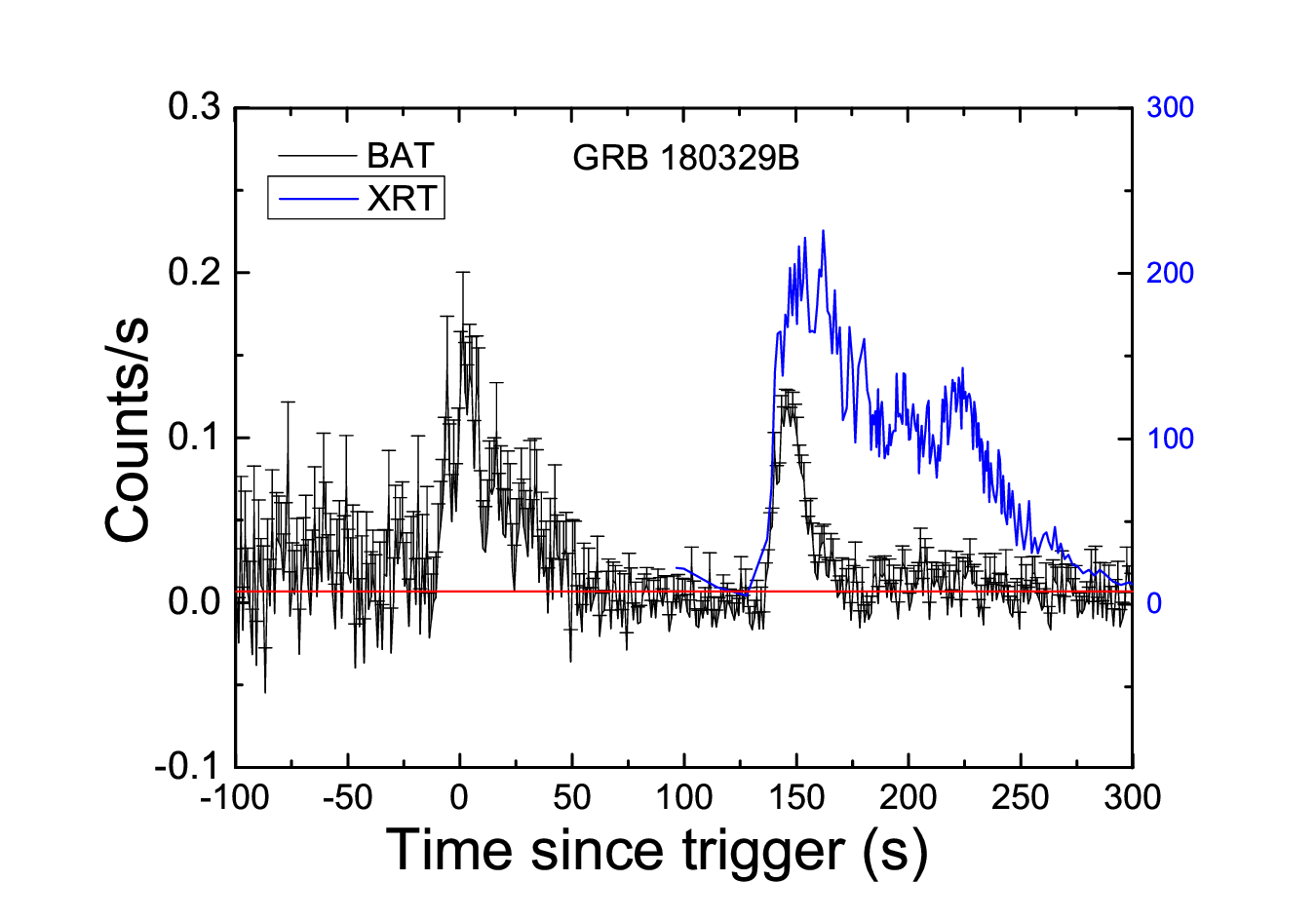}
\caption{BAT lightcurves (black) of prompt emission
and X-ray flares (blue) for our sample. The
solid red line is the background of prompt emission.}
\label{Fig2}
\end{figure*}

\begin{figure*}[h]
\centering
\includegraphics[angle=0,scale=0.25]{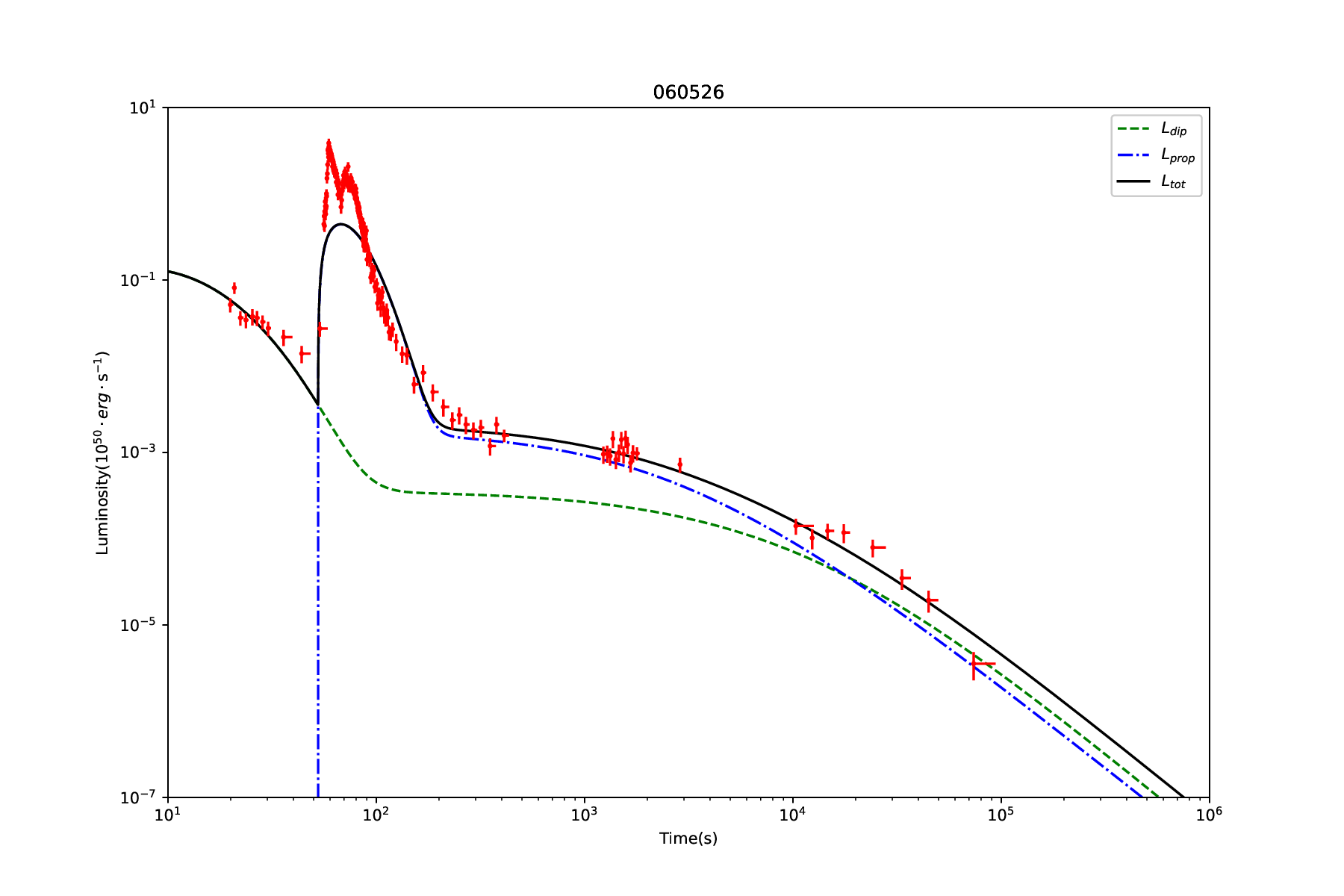}
\includegraphics[angle=0,scale=0.25]{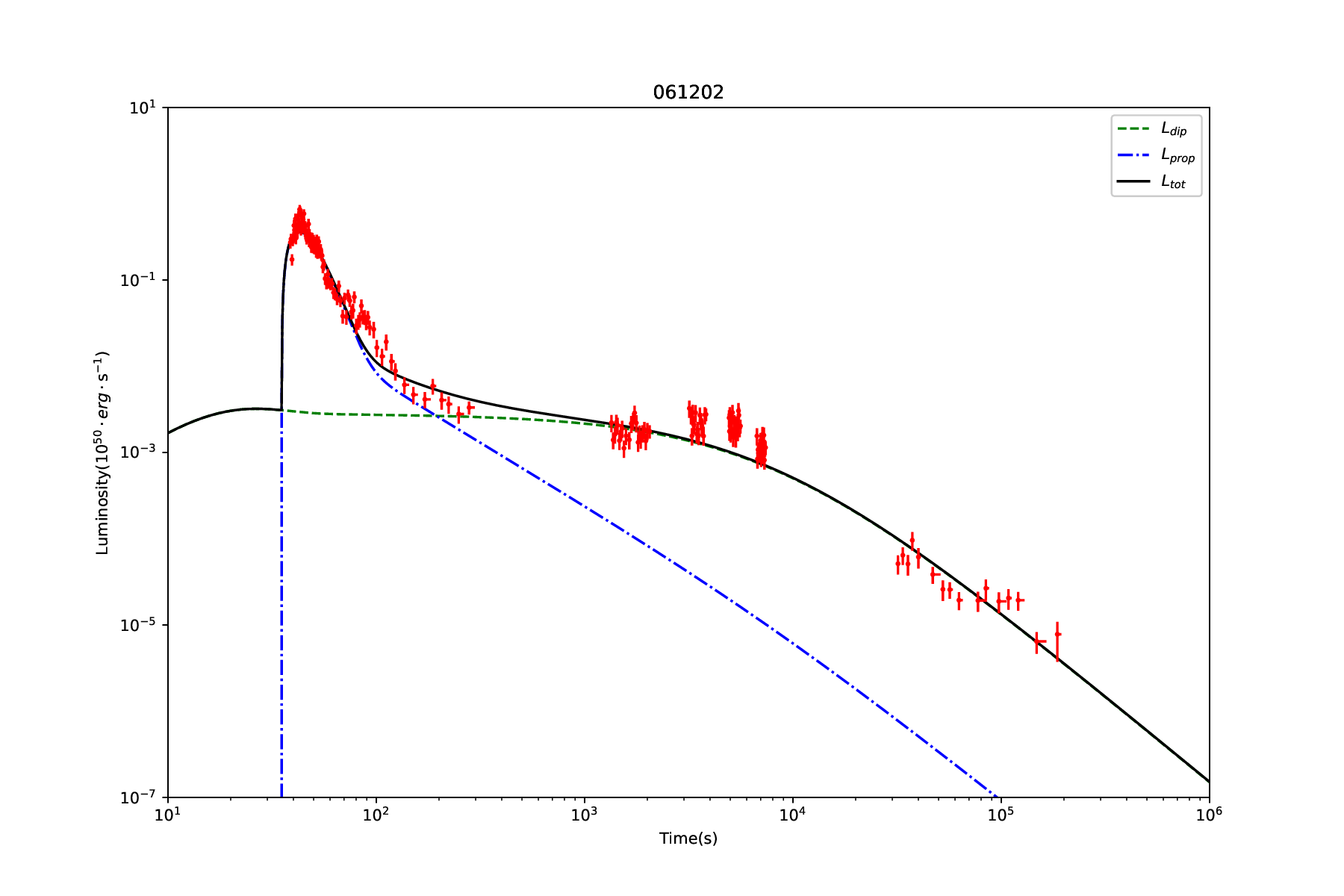}
\includegraphics[angle=0,scale=0.25]{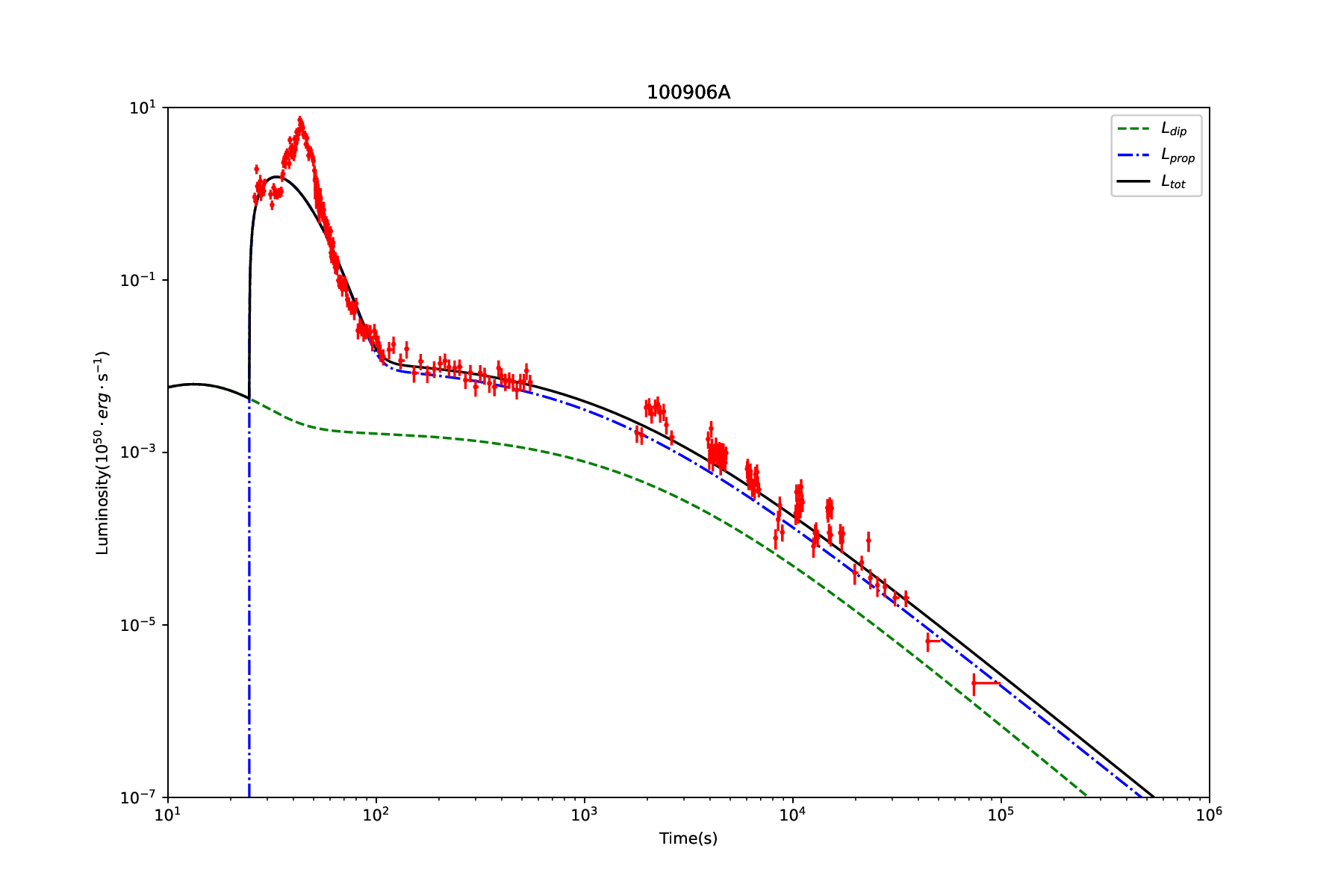}
\includegraphics[angle=0,scale=0.25]{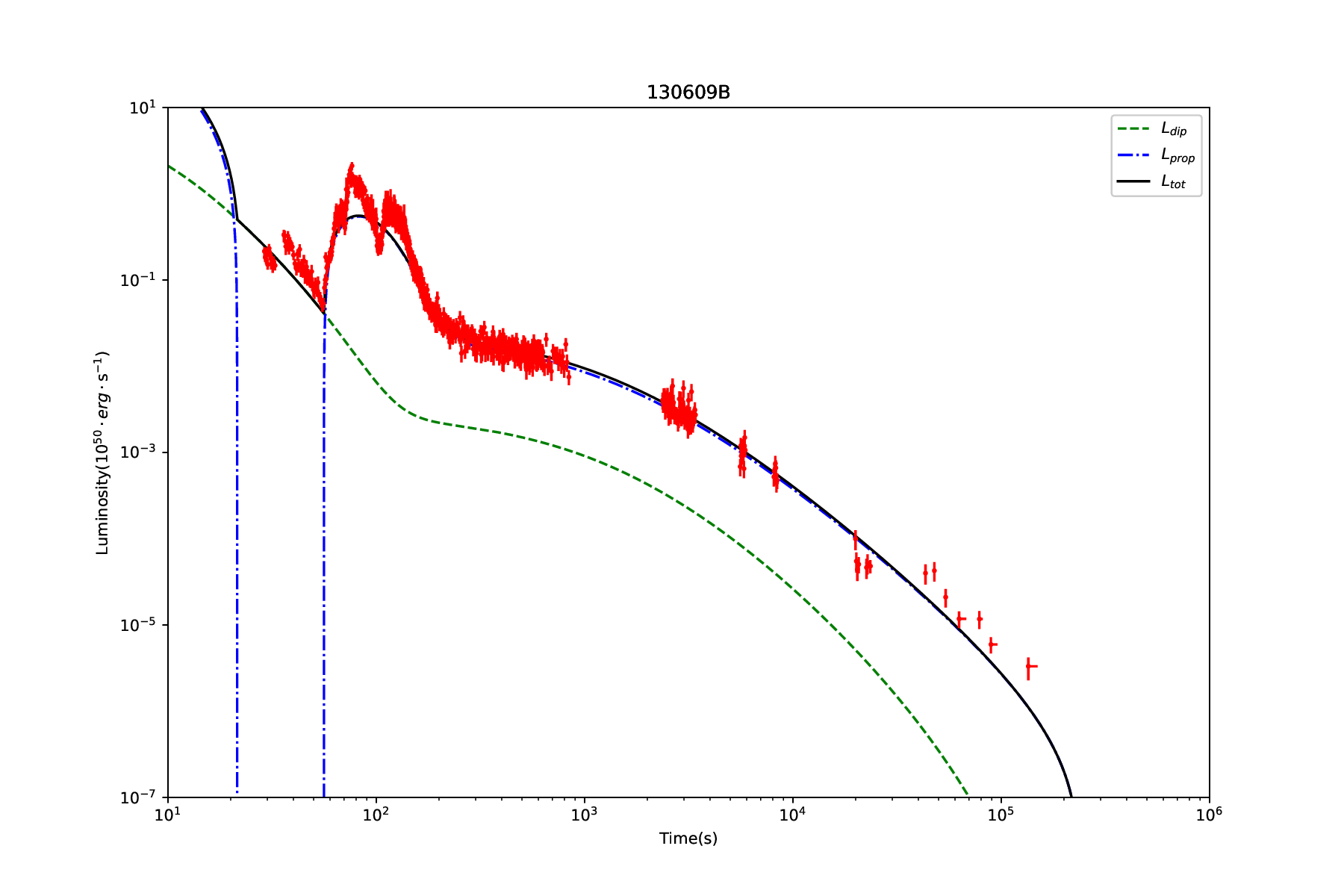}
\includegraphics[angle=0,scale=0.25]{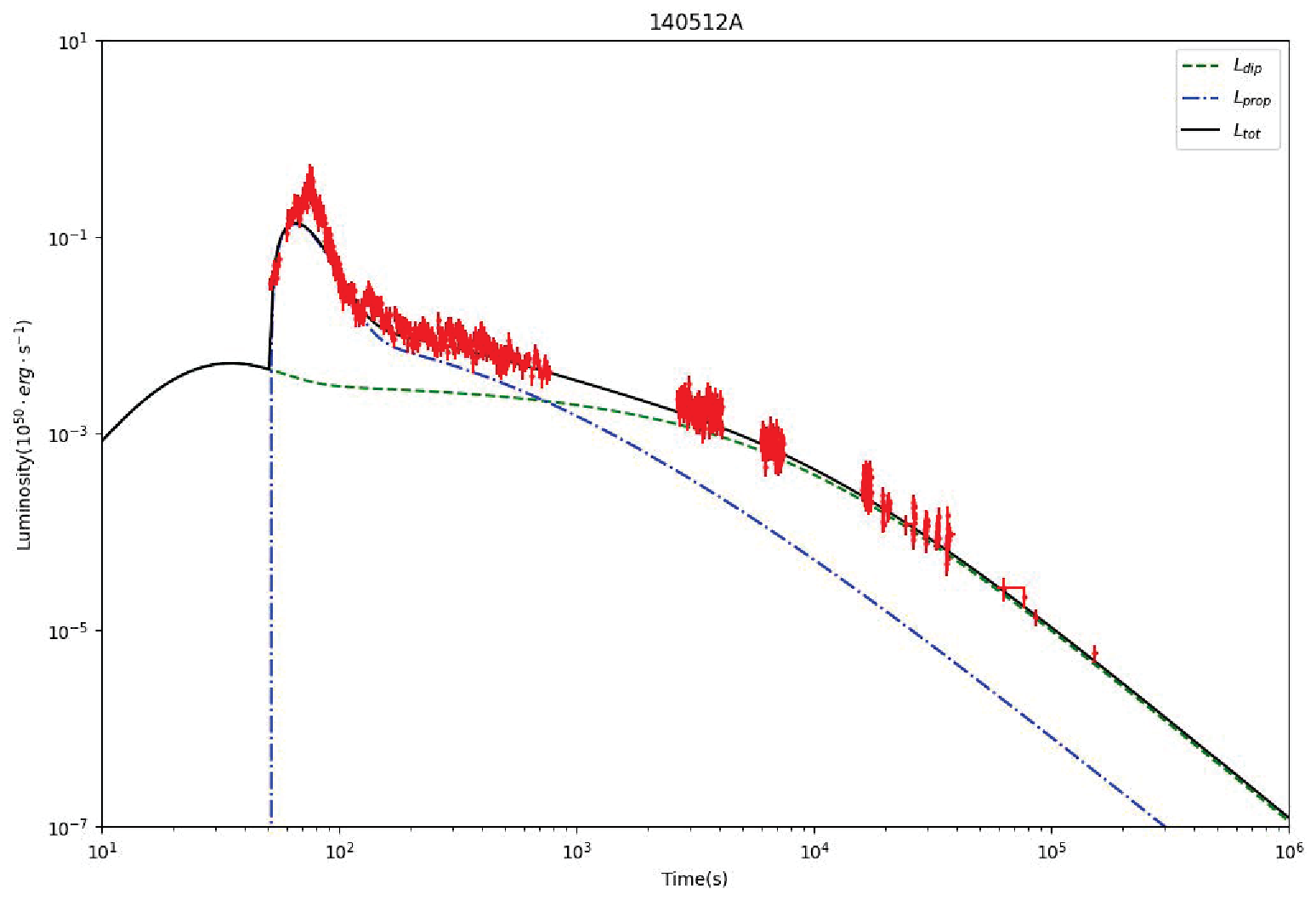}
\includegraphics[angle=0,scale=0.25]{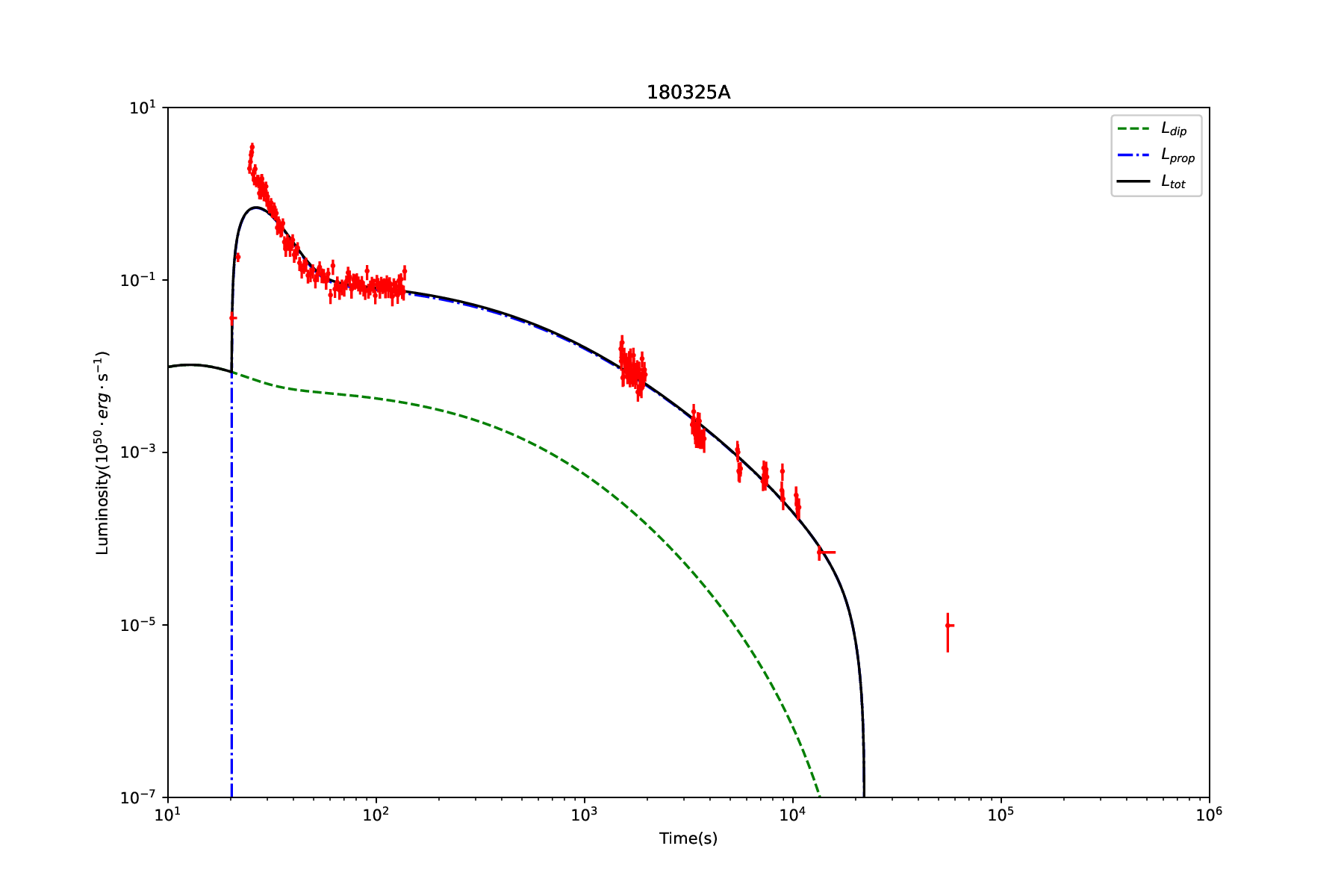}
\includegraphics[angle=0,scale=0.25]{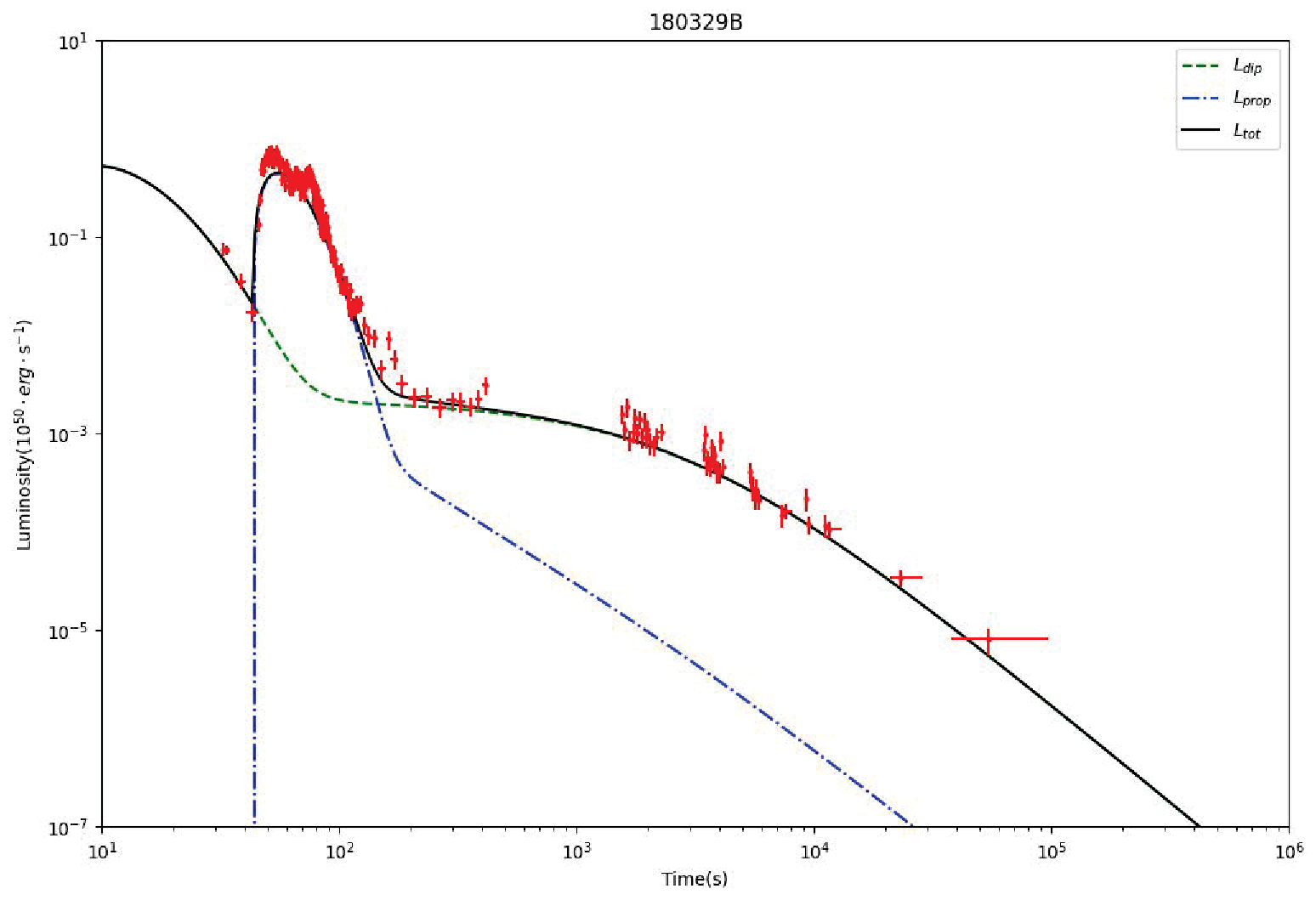}
%\hfill
\caption{X-ray light curves of GRBs for our sample (red points),
as well as the fits by magnetar propeller model with fallback accretion (red dash-dotted line)
and magnetar spin-down (blue dashed line). The
black solid line is the total luminosity of propeller and spin-down.}
\label{Fig3}
\end{figure*}

More interestingly, it is found that the X-ray flare/bump for those seven GRBs is simultaneously
observed by Swift/BAT in $\gamma-$ray band by searching in GCN Circulars Archive case by case. So
we downloaded the Swift/BAT data of the seven GRBs from the NASA Archive, and extracted the light
curves with the standard Swift scientific tools. Based on the methods from
\cite{2014ApJ...789..145H} for Swift/BAT, we employed the Bayesian Block (BBs;
\citealt{1998ApJ...504..405S}) algorithm to search for possible signals and identify the light
curves. In order to search for a low-significance signal before and after the duration, the
extracted BAT light curves usually cover 300 s before the BAT triggers and after duration of the
GRBs. More details, please refer to the paper on data analysis with the Bayesian Block algorithm
\citep{2014ApJ...789..145H}. Phenomenally, we find that the prompt emission in $\gamma-$ray band of
the seven GRBs are composed of two sub-bursts with the quiescent time. The light curves of the two
sub-bursts exhibit different behavior for our sample, e.g., a softer sub-burst prior to (3 out of
7) or following (4 out of 7) the stronger sub-burst of prompt emission. The light curves of both
prompt emission and X-ray flares for those seven GRBs are shown in Figure \ref{Fig2}.

%%%%%%%%%%%%%%%%%%%%%%%%%%%%%%%%%%%%%%%%%%%%%%%%%%%%%%
\section{Fitting results with magnetar propeller-fallback-accretion model}
One motivation is to test whether magnetar propeller-fallback-accretion model can be invoked to
interpret the GRBs that we selected sample above. In order to find out the best fit parameters of
magnetar propeller-fallback-accretion model, we adopt the MCMC simulation package
\citep{2013PASP..125..306F} to fit the X-ray flare/bump and X-ray plateau in the afterglow. There
are nine free parameters (i.e., $B_{\rm p}$, $P_{\rm 0}$, $M_{\rm d,0}$, $R_{\rm d}$, $\epsilon$,
$\delta$, $\eta_{\rm dip}$, $\eta_{\rm prop}$, and $1/f_{\rm B}$) by invoking the propeller and
dipole radiations of magnetar. The MCMC is one of the very popular methods in the field of high
energy astronomy, and it is very effective in constraining the free parameters which are in
degeneracy between each other. More details can refer to
\cite{2017MNRAS.470.4925G,2018MNRAS.478.4323G}.

Initially, we set the range of those parameters as following, $B_{\rm p}\sim(10^{-3}-10)\times
10^{15}$ G, $P_0\sim(0.5-100)\times 10^{-3}$ s, $M_{\rm d,0}\sim(10^{-3}-10^{2})M_{\odot}$, $R_{\rm
d}\sim(50-2000)\times 10^{5}$ cm, $\epsilon\sim(10^{-1}-10^{3})$, $\delta\sim(10^{-5}-50)$,
$\eta_{\rm dip}\sim(1-100)\%$, $\eta_{\rm prop}\sim(1-100)\%$, $1/f_B\sim(1-600)$. Then, we adopt
MCMC method to fit the X-ray light curve by invoking the propeller-fallback-accretion model and
spin-down dipole radiation of magnetar.

\begin{figure*}[h]
\centering
\includegraphics[angle=0,scale=0.15]{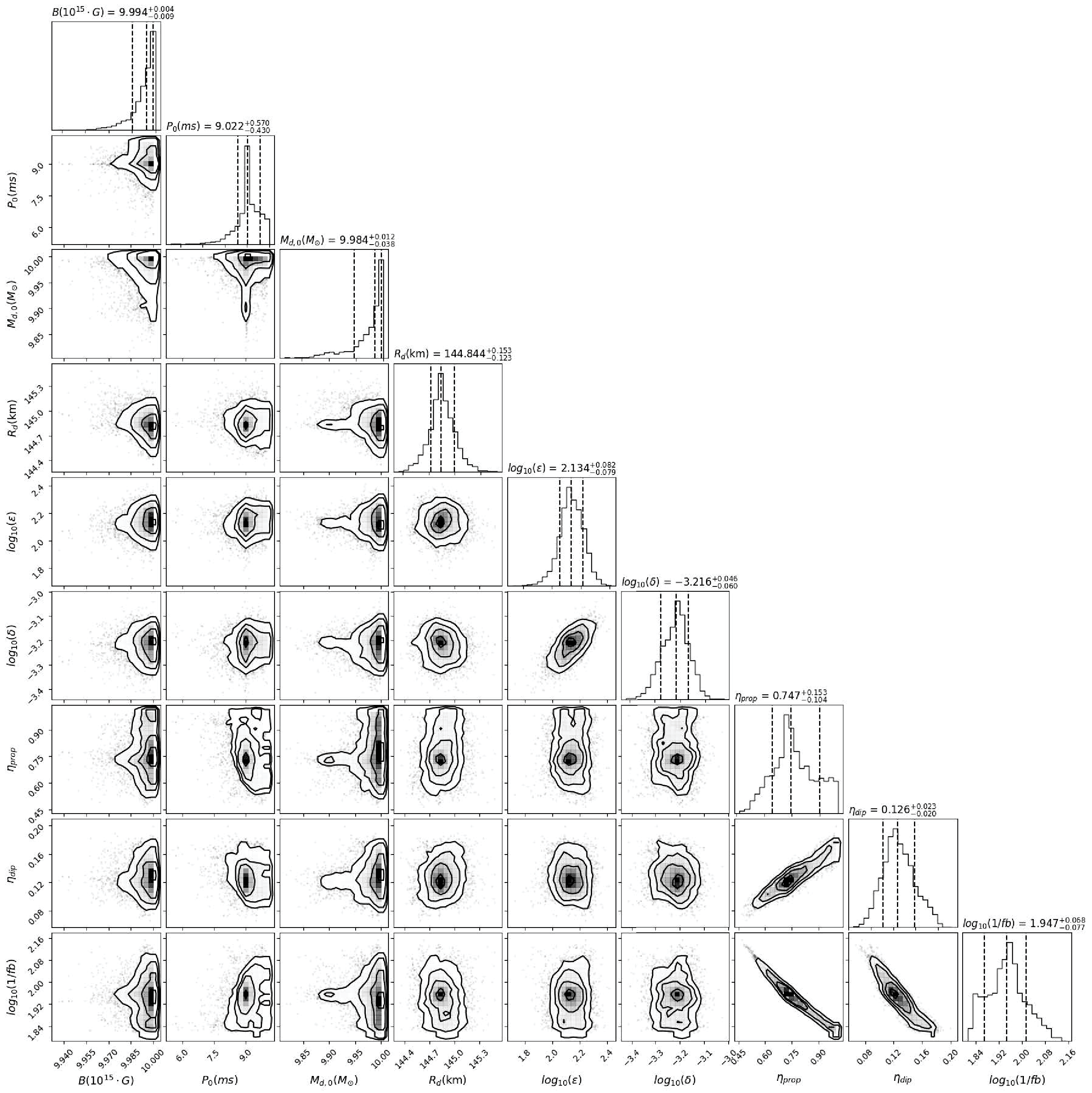}
\includegraphics[angle=0,scale=0.15]{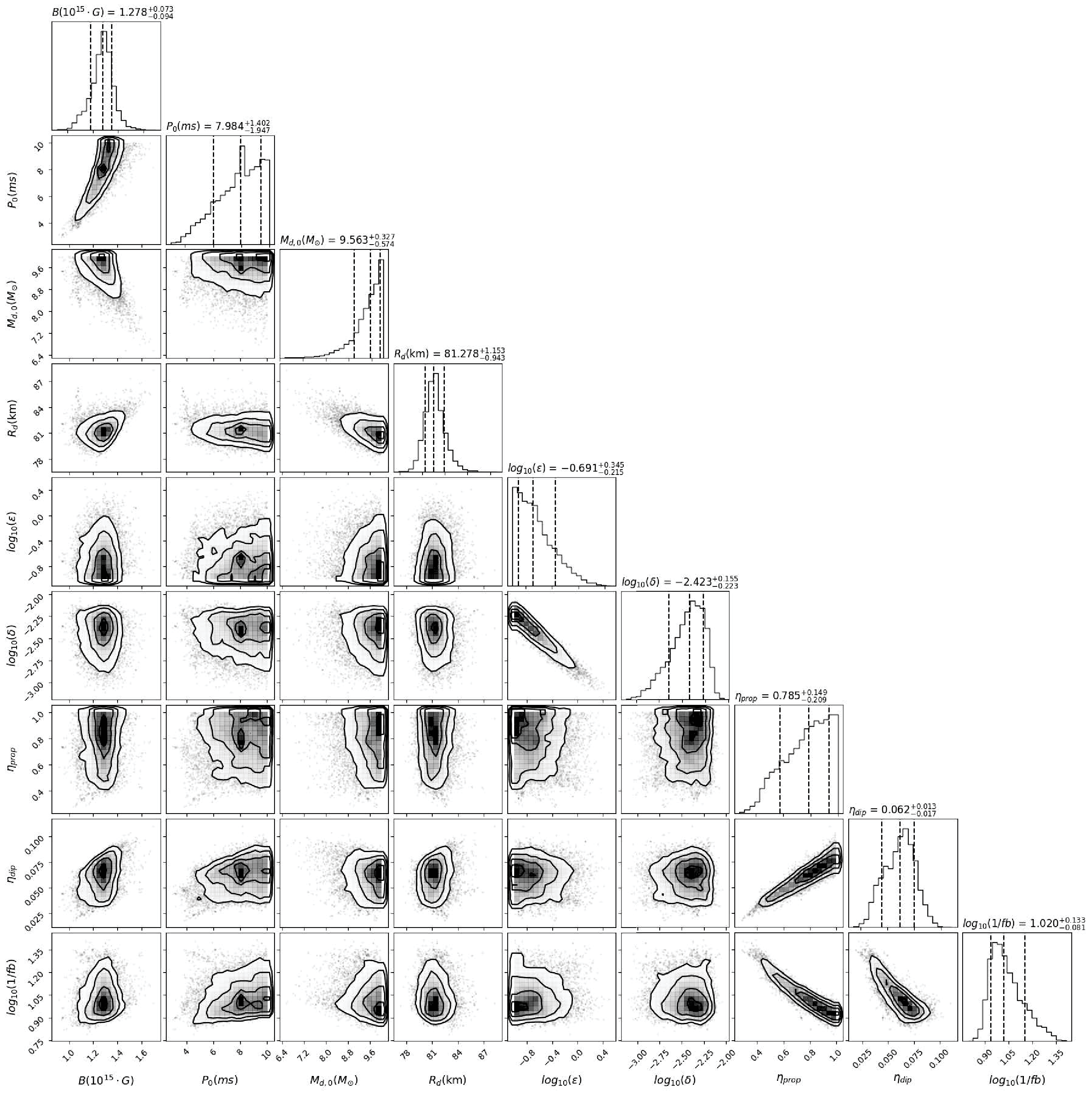}
\includegraphics[angle=0,scale=0.15]{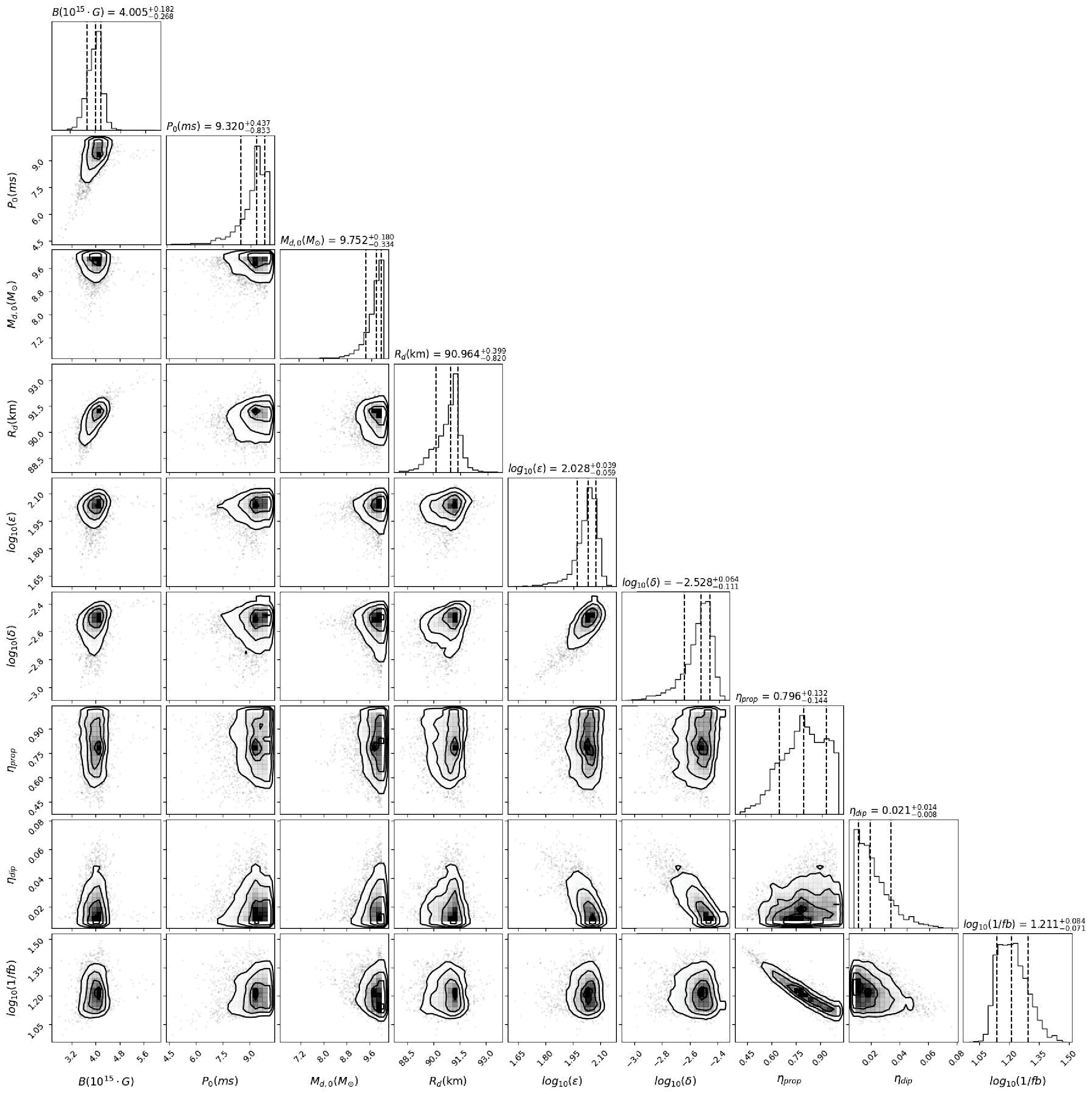}
\includegraphics[angle=0,scale=0.15]{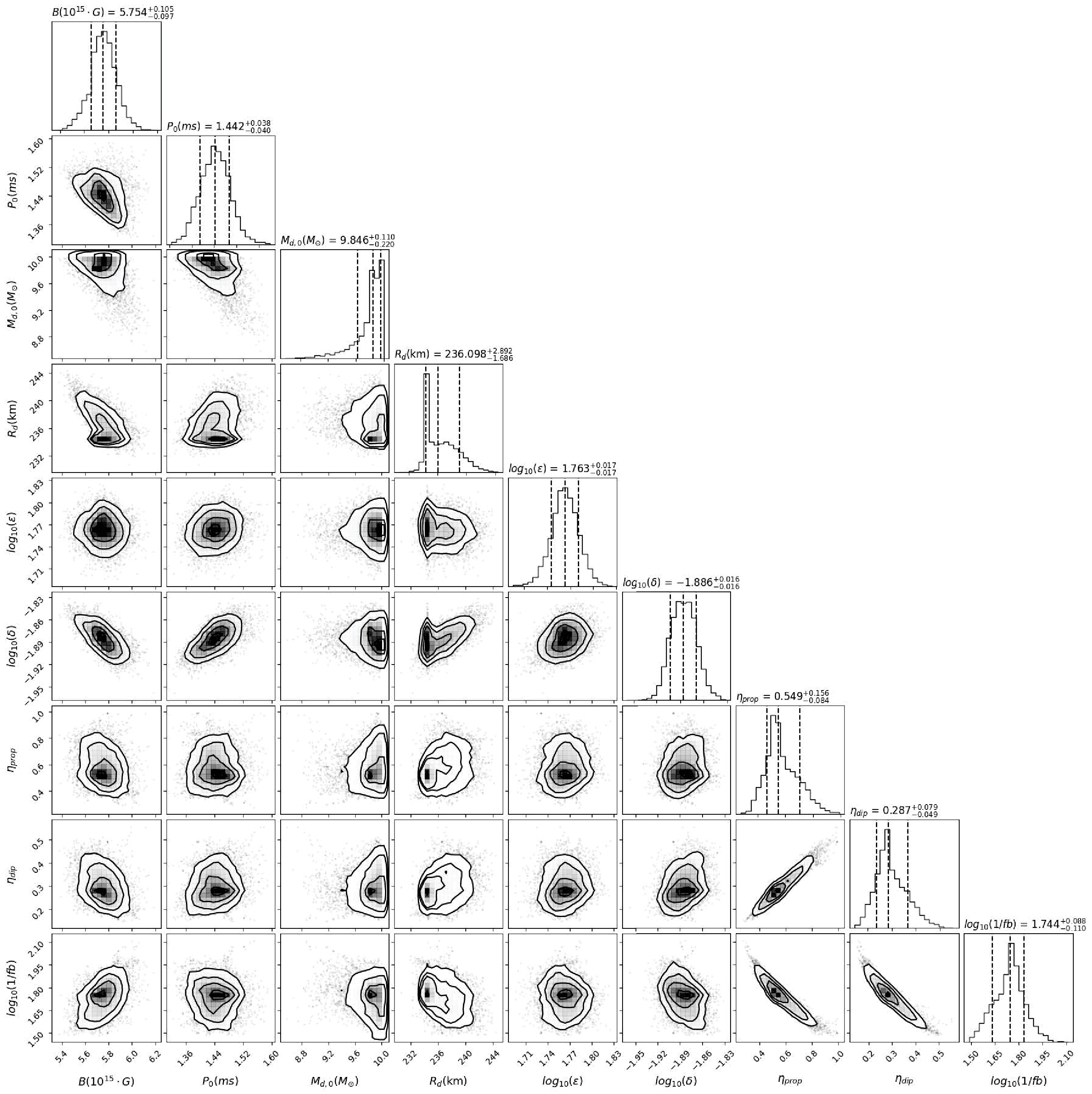}
\caption{2D histograms and parameter constraints of model fit for our sample. The 1D histograms
show the distribution for each parameter. The dashed lines indicate the median and $\pm 2\sigma$
uncertainty of the values.} \label{Fig4}
\end{figure*}

Figure \ref{Fig3} shows the MCMC fitting results by using the propeller-fallback-accretion model
and spin-down dipole radiation of magnetar for the seven GRBs. We find that the free parameters of
GRBs 140512A and 180329B can be constrained very well, such as $B_{\rm p}$ and $P_{0}$, which are
within a range of several times $10^{15}$ G and tens of milliseconds, respectively. The derived
parameters of $B_{\rm p}$, $P_{0}$, and $M_{\rm d,0}$ fall into the reasonable range, and they are
consistent with the results from \cite{2014ApJ...785...74L} and \cite{2018MNRAS.478.4323G}.
However, the free parameters of the other five cases are, more or less, not all constrained well
enough, especially, the main free parameters of propeller-fallback-accretion model, such as $B_{\rm
p}$, $P_{0}$, and $M_{\rm d,0}$. The two-dimensional histograms and parameter constraints of model
fit by invoking MCMC method for our sample are shown in Figure \ref{Fig4}, and the values of
parameter constraints are reported in Table 1. On the other hand, based on the fitting results, we
find that the required conversion efficiency of propeller is two or three times (or even large)
higher than that of spin-down dipole radiation of magnetar (see Figure \ref{Fig5} and Table 1).
Based on the results from \cite{2018ApJ...862..155L} for Fermi/GBM and \cite{2014ApJ...795..155P}
for Swift/BAT, it suggests that those two sub-bursts emission of GRB, together with the observed
simultaneous X-ray flare/bump, are likely to share the same physical origin. However, the higher
efficiency for the propeller phase in our fits is contradictory to the expectation from the
propeller model, namely, a dirtier ejecta should be less efficient in producing $\gamma-$ray
emissions. It means that at least the magnetar propeller-fallback-accretion model cannot interpret
very well to the GRB X-ray flares that are simultaneously observed in $\gamma-$ray band, even
though it can be invoked to interpret the X-ray plateaus for short GRBs \citep{2014MNRAS.438..240G}
and supernova explosion \citep{2011ApJ...736..108P}.
%%%%%%%%%%%%%%%%%%%%%%%%%%%%%%%%%%%%%%%%%%%%%%%%%%%%

\section{Conclusion and Discussion}
The prompt emission, X-ray plateau, and X-ray flares of GRBs are thought to be from internal
dissipation \citep{2011CRPhy..12..206Z}. The magnetar as the central engine of GRB has been
proposed to interpret the observed X-ray flares and plateau emission
\citep{2014MNRAS.438..240G,2017MNRAS.470.4925G}. By systematically searching for X-ray emission
observed by Swift/XRT, we find that seven robust GRBs include both X-ray flares and plateau
emissions with redshift measured. More interestingly, it is found that the X-ray flare/bump for
those seven GRBs is simultaneously observed by Swift/BAT in $\gamma-$ray band. The prompt emission
in $\gamma-$ray band of the seven GRBs is composed of two sub-bursts with the quiescent time.
Especially, the second sub-burst emission in the prompt emission is observed simultaneously with
X-ray flare.

Then, by adopting the MCMC method, we invoke the propeller-fallback-accretion model and spin-down
dipole radiation of magnetar central to fit the X-ray data of plateau emissions and flares for our
sample. It is found that the free parameters of two GRBs (GRBs 140512A and 180329B) can be
constrained very well, such as $B_{\rm p}$ and $P_{0}$ are range of several times $10^{15}$ G and
tens of milliseconds, respectively. However, the free parameters of the other five cases, more or
less, are not all constrained well enough, especially, the main free parameters of
propeller-fallback-accretion model, such as $B_{\rm p}$, $P_{0}$, and $M_{\rm d,0}$. On the other
hand, it requires that the conversion efficiency of propeller is two or three times higher than
that of spin-down dipole radiation of magnetars. If this is the case, the higher
efficiency for the propeller phase in our fits is contradictory to the expectation from the
propeller model, namely, a dirtier ejecta should be less efficient in producing $\gamma-$ray
emissions. Our results suggest that at least the magnetar propeller-fallback-accretion model cannot
interpret very well to the GRB X-ray flares that are simultaneously observed in $\gamma-$ray.

We believe that the magnetar propeller-fallback-accretion model can interpret very well the X-ray
plateau emissions of some GRBs. However, this model exists significant challenges and flaws for the
baryon loading issue when it is related to $\gamma-$ray emission and high relativistic outflow.
From the observational point of view, the X-ray flares are observed in a good fraction of GRB
afterglow, and the constrained bulk Lorentz factor (or just upper limit) of the X-ray flare
outflows ranges from a few tens to hundreds \citep{2010ApJ...724..861J,2015ApJ...807...92Y}. It
means that the dirty of baryon loading issue still exists if we invoke the magnetar
propeller-fallback-accretion model to interpret the X-ray flares with a highly Lorentz factor.
Furthermore, several physical models without magnetar central engine are invoked to interpret the
observed X-ray flare, plateau emission, as well as two sub-bursts $\gamma-$ray emission, such as
late central engine reactivity \citep{2005Sci...309.1833B,2005MNRAS.364L..42F,2006Sci...311.1127D},
the relativistic jet and cocoon emissions
\citep{2002MNRAS.337.1349R,2010ApJ...725.1137L,2017ApJ...834...28N}, two steps in the collapse of
the progenitor star \citep{2009MNRAS.397.1695L}, the collapse of a rapidly rotating stellar core
leading to fragmentation \citep{2005ApJ...630L.113K}, fragmentation in the accretion disc
\citep{2006ApJ...636L..29P}, the magnetic barrier around the accretor \citep{2006MNRAS.370L..61P},
the gravitational lensing \citep{2021NatAs...5..560P,2021ApJ...921L..29Y,2022ApJ...931....4L}, and
jet precession model \citep{2023ApJ...945...17G}. However, more or less, each model above cannot
fully interpret all properties of observations for two sub-bursts emission of GRB from light curve,
spectrum, as well as the quiescent time.

\begin{acknowledgements}
We acknowledge the use of the public data from the UK Swift Science Data Centre. This work is
supported by the Guangxi Science Foundation (grant Nos. 2023GXNSFDA026007 and 2017GXNSFFA198008),
the National Natural Science Foundation of China (grant No. 11922301), and the Program of Bagui
Scholars Program (LHJ). Lin Lan is supported by the National Postdoctoral Program for Innovative
Talents (grant No. GZB20230765).
\end{acknowledgements}

%*********************************************reference********************************************
%\clearpage

%****************************************end***************************************************

%%%%%%%%%%%%%%%%%%%%%%%%%%%%%%%%%%%%%%%%%%%%%%%%%%%%
\begin{sidewaystable}[!htp]
\begin{threeparttable}
    \centering
    \caption{The values of derived parameters from the MCMC for our sample.}
    \renewcommand{\arraystretch}{3}
    \begin{tabular}{c c c c c c c c c c c}
         \hline
 GRB & \makecell[c]{$B_{\rm p}$\\$(10^{15} \cdot  \rm G)$} & \makecell[c]{$P_0$\\$\rm (ms)$} &
 \makecell[c]{$M_{\rm d,0}$\\$(10^{-2} \cdot M_{\odot})$} & \makecell[c]{$R_{\rm d}$\\$\rm (km)$} & $
 \epsilon$ & \makecell[c] {$\delta$ \\$10^{-4}$} & $\eta_{\rm prop}$ & $\eta_{\rm dip}$ & $ 1/f_B$ & $ z $\\
\hline

140512A & $2.28^{+0.01}_{-0.01}$ & $69.85^{+1.51}_{-0.82}$ & $8.13^{+0.04}_{-0.04}$ &
$148.95^{+0.14}_{-0.12}$ & $19.72^{+0.23}_{-0.23}$ & $73.62^{+0.34}_{-0.34}$ & $0.56^{+0.01}_{-0.01}$ &
$0.23^{+0.01}_{-0.01}$ & $7.93^{+0.13}_{-0.14}$ & $0.725^{(a)}$\\

180329B & $6.30^{+0.01}_{-0.01}$ & $29.93^{+0.23}_{-0.24}$ & $0.44^{+0.01}_{-0.01}$ &
$116.53^{+0.10}_{-0.10}$ & $0.28^{+0.02}_{-0.02}$ & $2.32^{+0.12}_{-0.11}$ & $0.24^{+0.01}_{-0.01}$ &
$0.08^{+0.01}_{-0.01}$ & $29.04^{+1.30}_{-1.05}$& $1.998^{(b)}$\\\

060526 & $9.99^{+0.01}_{-0.01}$ & $9.02^{+0.57}_{-0.43}$ & $9.98^{+0.01}_{-0.03}$ & $144.84^{+0.15}_{-0.12}$
& $136.14^{+9.06}_{-22.64}$ & $6.08^{+0.67}_{-0.78}$ & $0.74^{+0.15}_{-0.10}$ & $0.12^{+0.02}_{-0.02}$ &
$88.51^{+15.00}_{-14.38}$& $3.21^{(c)}$\\

061602 & $1.278^{+0.07}_{-0.09}$ & $7.98^{+1.40}_{-1.94}$ & $9.56^{+0.32}_{-0.57}$ & $81.27^{+1.15}_{-0.94}$
& $0.20^{+0.24}_{-0.08}$ & $37.75^{+16.19}_{-15.16}$ & $0.78^{+0.15}_{-0.21}$ & $0.06^{+0.01}_{-0.02}$ &
$10.47^{+3.75}_{-1.78}$ & $2.253^{(d)}$\\

100906A & $4.00^{+0.18}_{-0.26}$ & $9.32^{+0.43}_{-0.83}$ & $9.75^{+0.18}_{-0.33}$ & $90.96^{+0.39}_{-0.82}$
& $161.43^{+15.16}_{-20.50}$ & $29.64^{+4.70}_{-6.68}$ & $0.79^{+0.13}_{-0.14}$ & $0.02^{+0.01}_{-0.01}$ &
$16.25^{+3.46}_{-2.45}$ & $1.727^{(e)}$\\

130609B & $5.75^{+0.10}_{-0.09}$ & $1.44^{+0.03}_{-0.04}$ & $9.84^{+0.11}_{-0.22}$ &
$236.09^{+2.89}_{-1.68}$
& $57.94^{+2.31}_{-2.22}$ & $130.01^{+4.87}_{-4.70}$ & $0.55^{+0.15}_{-0.08}$ & $0.28^{+0.08}_{-0.05}$ &
$55.46^{+12.45}_{-12.40}$ & $1.3^{(f)}$\\

180325A & $3.29^{+0.09}_{-0.07}$ & $9.68^{+0.21}_{-0.45}$ & $9.77^{+0.15}_{-0.36}$ & $65.67^{+1.37}_{-0.95}$
& $109.14^{+6.46}_{-6.10}$ & $883.07^{+28.93}_{-35.85}$ & $0.72^{+0.17}_{-0.16}$ & $0.08^{+0.14}_{-0.06}$ &
$6.35^{+1.94}_{-1.10}$ & $2.25^{(g)}$\\

\hline
\end{tabular}
 \begin{tablenotes}
 \footnotesize
\item[]References of redshift measured: (a) \cite{2014GCN.16310....1D}; (b) \cite{2018GCN.22567....1I};
(c) \cite{2006GCN..5170....1B}; (d) \cite{2016ApJ...817....7P}; (e) \cite{2010GCN.11230....1T}; (f)
\cite{2013GCN.14888....1R}; (g) \cite{2018GCN.22535....1H}.
 \end{tablenotes}
\end{threeparttable}
    \label{fitting results}
\end{sidewaystable}

%%%%%%%%%%%%%%%%%%%%%%%%%%%%%%%%%%%%%%%%%%%%%%%%%%%%

%%%%%%%%%%%%%%%%%%%%%%%%%%%%%%%%%%%%%%%%%%%%%%%%%%%%%%%%%%%%%%%%%%%%%%%%%%%%%%%%%
\clearpage

%%%%%%%%%%%%%%%%%%%%%%%%%%%%%%%%%%%%%%%%%%%%%%%%%%%%%%%%%%%%%%%%%%%%%%%%%%%%%%%%%

%%%%%%%%%%%%%%%%%%%%%%%%%%%%%%%%%%%%%%%%%%%%%%%%%%%%%%%%%%%%%%%%%%%%%%%%%%%%%%%%%

%%%%%%%%%%%%%%%%%%%%%%%%%%%%%%%%%%%%%%%%%%%%%%%%%%%%%%%%%%%%%%%%%%%%%%%%%%%%%%%%%

%%%%%%%%%%%%%%%%%%%%%%%%%%%%%%%%%%%%%%%%%%%%%%%%%%%%%%%%%%%%%%%%%%%%%%%%%%%%%%%%%
\begin{figure*}[h]
\centering
\includegraphics[angle=0,scale=0.15]{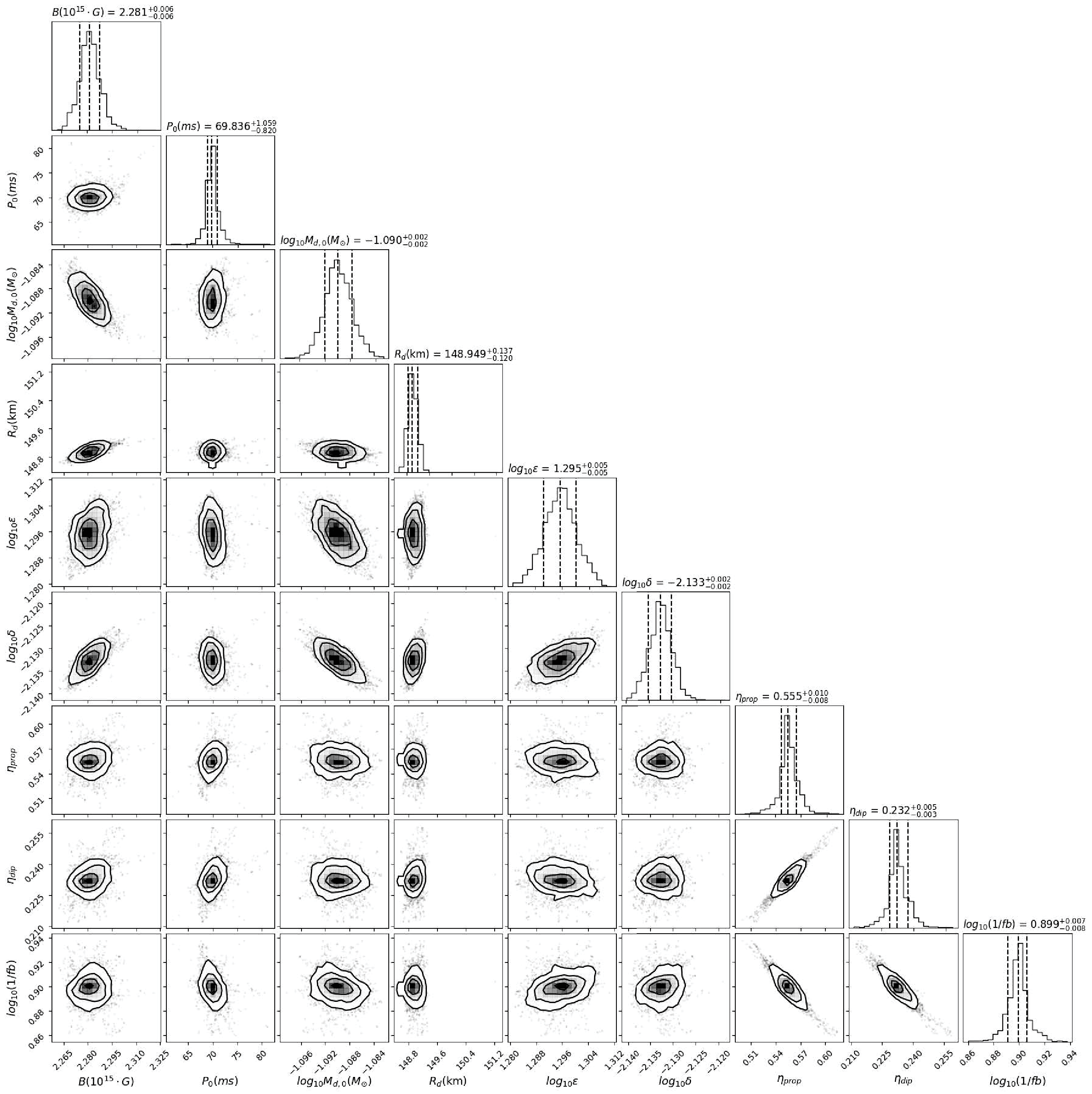}
\includegraphics[angle=0,scale=0.15]{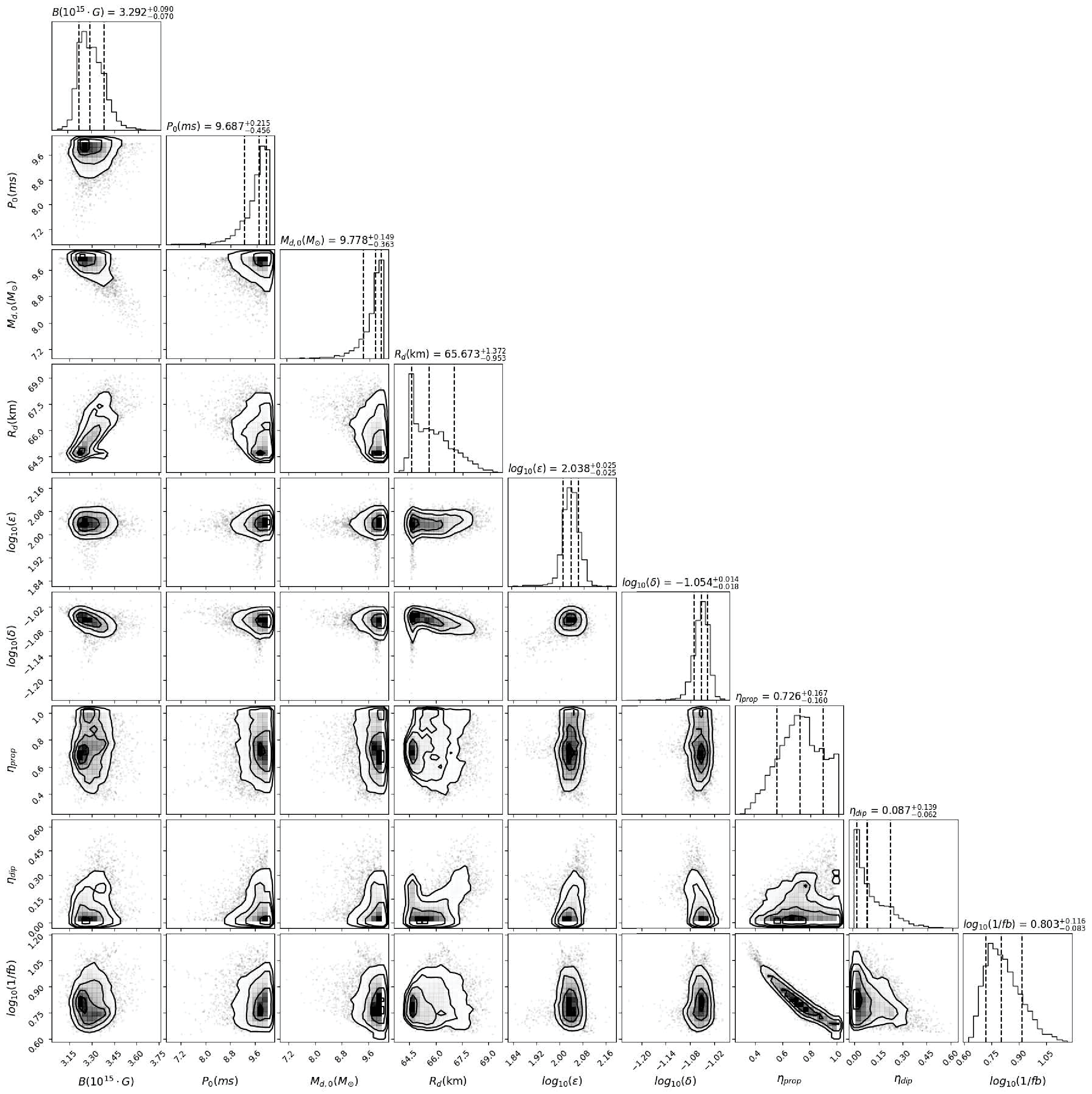}
\includegraphics[angle=0,scale=0.15]{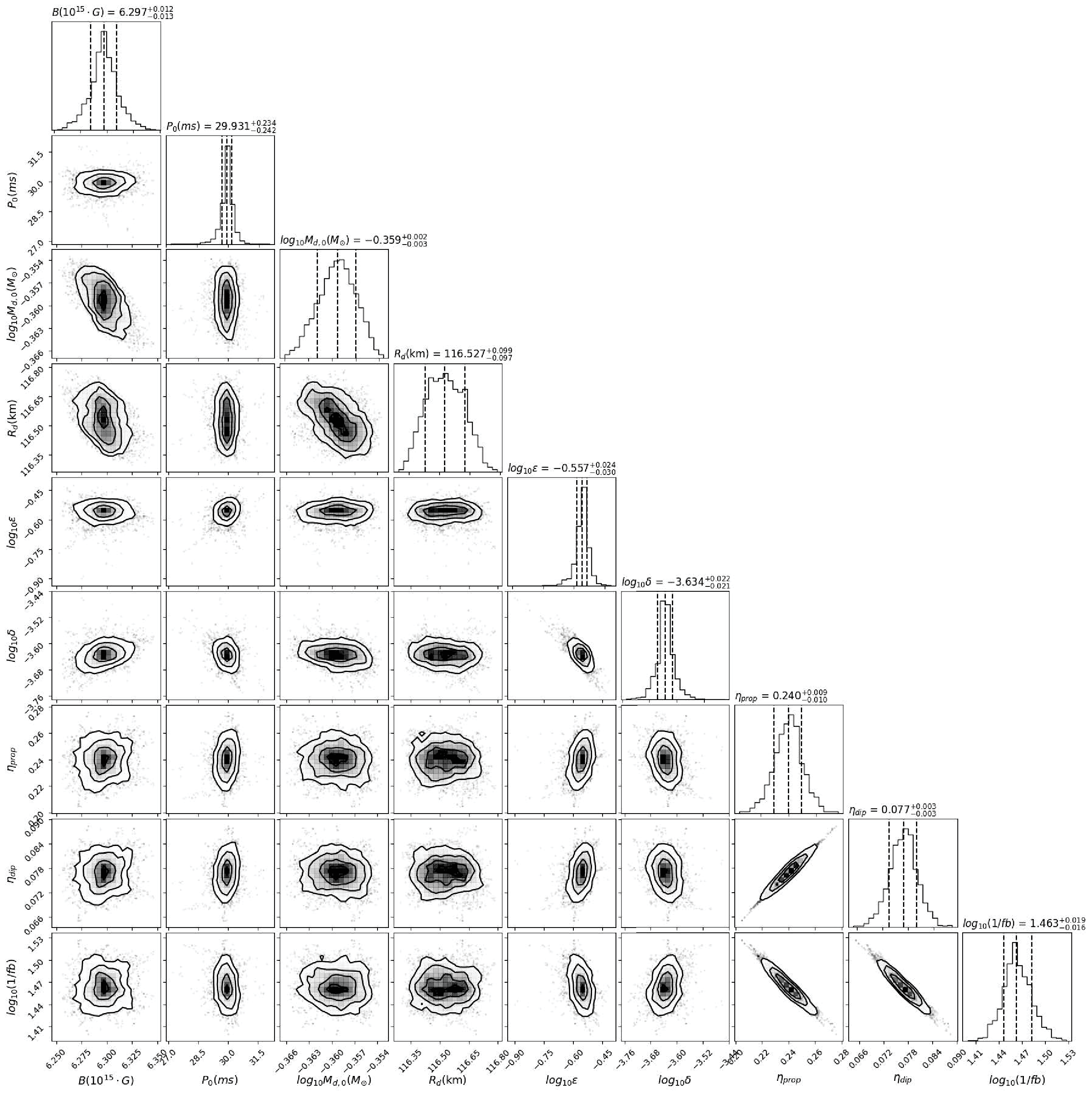}
\center{Fig. 4---  continued.}
\end{figure*}
%%%%%%%%%%%%%%%%%%%%%%%%%%%%%%%%%%%%%%%%%%%%%%%%%%%%%%%%%%%%%%%%%%%%%%%%%%%%%%%%%
\begin{figure*}[h]
\centering
\includegraphics[angle=0,scale=0.6]{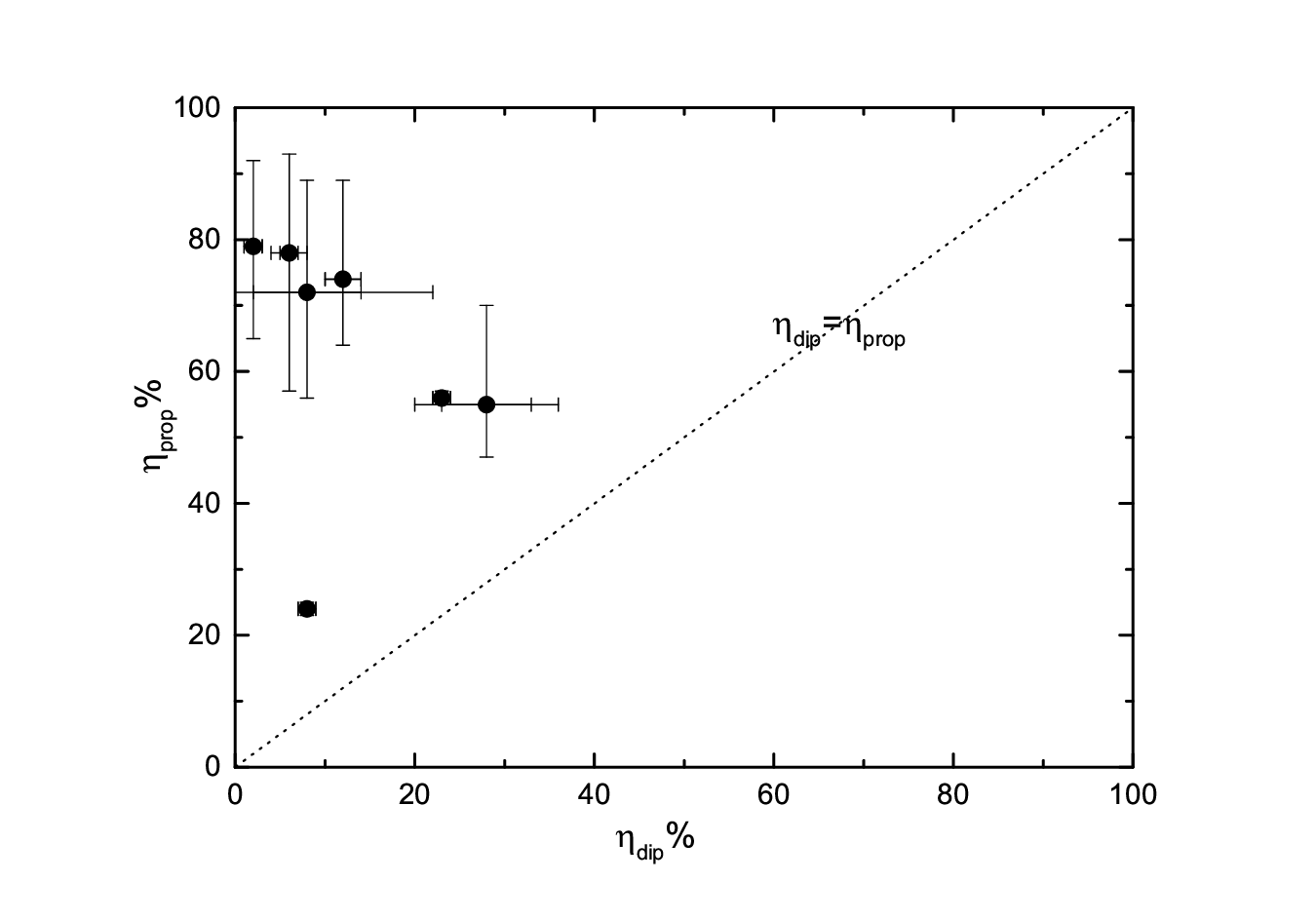}
\caption{Efficiency correlation between propeller and dipole radiation. The dotted line corresponds
to $\eta_{\rm prop}=\eta_{\rm dip}$.} \label{Fig5}
\end{figure*}

%%%%%%%%%%%%%%%%%%%%%%%%%%%%%%%%%%%%%%%%%%%%%%%%%%%%%%%%%%%%%%%%%%%%%%%%%%%%%%%%%
%%%%%%%%%%%%%%%%%%%%%%%%%%%%%%%%%%%%%%%%%%%%%%%%%%%%%%%%%%%%%%%%%%%%%%%%%%%%%%%%%
\end{document}